\documentclass[sigconf]{acmart}
\AtBeginDocument{%
  \providecommand\BibTeX{{%
    \normalfont B\kern-0.5em{\scshape i\kern-0.25em b}\kern-0.8em\TeX}}}

\setcopyright{acmcopyright}
\copyrightyear{2018}
\acmYear{2018}
\acmDOI{XXXXXXX.XXXXXXX}

\acmPrice{15.00}
\acmISBN{978-1-4503-XXXX-X/18/06}

\usepackage{subfigure}
\usepackage{algorithm} 
\usepackage{algorithmic} 

\usepackage{multirow} 
\usepackage{enumitem} 
\usepackage{amsmath}
\usepackage{xcolor}
\usepackage{subfigure}
\usepackage{mwe}
\usepackage{booktabs}
\usepackage{threeparttable}
\usepackage{algorithm}
\usepackage{algorithmic}
\usepackage{amsmath}
\usepackage{amsfonts}
\usepackage[switch]{lineno}

\begin{document}

\title{Cross-domain recommendation via user interest alignment}

\author{Chuang Zhao}
\affiliation{%
  \institution{College of Management and Economics, Tianjin University}
  \city{Tianjin}
  \country{China}}
\email{zhaochuang@tju.edu.cn}

\author{Hongke Zhao}
\affiliation{%
  \institution{College of Management and Economics, Tianjin University}
  \city{Tianjin}
  \country{China}}
\email{hongke@tju.edu.cn}

\author{Ming HE}
\affiliation{%
  \institution{AI Lab at Lenovo Research}
  \city{Beijin}
  \country{China}}
\email{heming01@foxmail.com}

\author{Jian Zhang}
\affiliation{%
  \institution{School of Cyberspace Security, Hangzhou Dianzi University}
  \city{Hangzhou}
  \country{China}}
\email{zhang.jian-94@outlook.com}

\author{Jianping Fan}
\affiliation{%
  \institution{AI Lab at Lenovo Research}
  \city{Beijin}
  \country{China}}
\email{jfan1@lenovo.com}

\renewcommand{\shortauthors}{Trovato and Tobin, et al.}

\begin{abstract}
Cross-domain recommendation aims to leverage knowledge from multiple domains to alleviate the data sparsity and cold-start problems in traditional recommender systems. One popular paradigm is to employ overlapping user representations to establish domain connections, thereby improving recommendation performance in all scenarios. Nevertheless, the general practice of this approach is to train user embeddings in each domain separately and then aggregate them in a plain manner, often ignoring potential cross-domain similarities between users and items. 
Furthermore, considering that their training objective is recommendation task-oriented without specific regularizations, the optimized embeddings disregard the interest alignment among user's views, and even violate the user's original interest distribution.
To address these challenges, we propose a novel cross-domain recommendation framework, namely \textit{COAST}, to improve recommendation performance on dual domains by perceiving the cross-domain similarity between entities and aligning user interests.
Specifically, we first construct a unified cross-domain heterogeneous graph and redefine the message passing mechanism of graph convolutional networks to capture high-order similarity of users and items across domains. Targeted at user interest alignment, we develop deep insights from two more fine-grained perspectives of user-user and user-item interest invariance across domains 
by virtue of affluent unsupervised and semantic signals.
We conduct intensive experiments on multiple tasks, constructed from two large recommendation data sets. Extensive results show \textit{COAST} consistently and significantly outperforms state-of-the-art cross-domain recommendation algorithms as well as classic single-domain recommendation methods. 
\end{abstract}


\maketitle

\begin{figure}[!ht]
\centering
\includegraphics[width=\linewidth,height=0.55\linewidth]{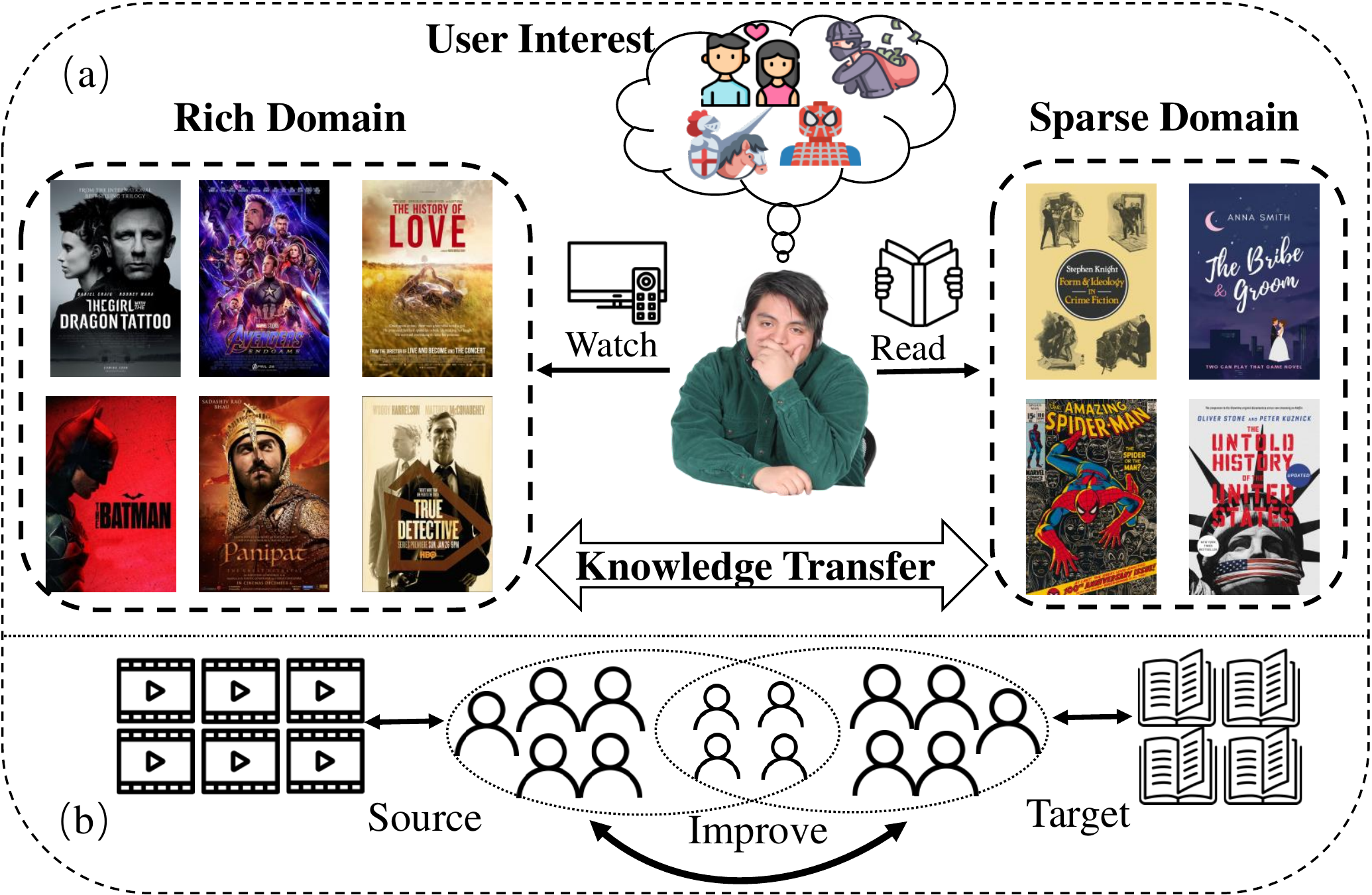}
\centering
\caption{(a) illustrates that the behavior of an overlapping user in different domains is driven by the same distribution of interests.  (b) illustrates that we improve the recommendation performance of the two domains through knowledge transfer of overlapping users.
}
\label{photo}
\end{figure}


\section{Introduction}
In an effort to alleviate information overload~\cite{naumov2019deep,da2020recommendation}, various well-known platforms such as Netflix~\cite{gomez2015netflix} and Amazon~\cite{linden2003amazon} deploy recommender systems to capture users' personalized preferences. Despite their excellent performance, data sparsity and cold-start problems, as two serious challenges, pose obstacles to model user interests accurately and efficiently~\cite{zhang2019deep}.

To address these headaches, researchers put their insights into cross-domain recommender systems (CDR), i.e., transfer knowledge from informative recommendation scenarios (source domain) to scenarios with sparse interactions (target domain) via transfer learning techniques~\cite{zang2021survey}. This directed transfer essentially enhances the knowledge of the target domain and achieves promising results on multiple recommendation data sets~\cite{zhu2022personalized}. Further, several researchers engage in bidirectional cross-domain recommendation, arguing that reasonable model structures can facilitate the mutual transfer of source and target domain knowledge~\cite{li2021dual}. For instance, user \emph{Jack} searches and browses a large number of computer cost-effective related posts in the online community (source domain), and we can simultaneously recommend various types of computers to him in the online mall (target domain), and vice versa. This dual recommendation paradigm can not only alleviate the negative transfer phenomenon, but also promote the upper bound on the target domain by improving the recommendation ability of the model in the source domain~\cite{zhu2021cross}. 

To our best knowledge, the mainstream taxonomy of dual cross-domain recommendation can be separated into \textit{collective matrix factorization, mapping-based methods, graph neural network-based approaches}, and \textit{representation combination of overlapping entities}~\cite{zang2021survey}. This paper strives to kick the last paradigm upstairs, the general practice of which is to train user and item representations separately in the two domains, and then perform specific aggregations (concat, dot, pooling) on them for knowledge transfer~\cite{zhu2019dtcdr}.
Even with the remarkable results~\cite{zhao2020catn}, they still 
encounter three serious challenges. Firstly, vast majority of these studies conduct experiments on explicit data sets with fully overlapping users, which significantly 
pole apart from rich implicit content and partial user overlap in real-world scenarios~\cite{man2017cross}. Secondly, the general practice of independently training entity representations in each domain structurally isolates the interactions among users-items, thereby failing to perceive higher-order similarities between entities. Thirdly, considering the recommendation task-oriented optimization objective, these work cannot guarantee the alignment of overlapping users' interests across domains~\cite{cao2022cross}. In other words, we argue that the plain aggregations of entity representations across domains without any regularizations are incapable of distinguishing users' personal preferences at the instance level, nor can it ensure that users' interests in items are consistent, or even cause conflicting interests among users' views.

With the aim of addressing these challenges, we propose a \textit{\textbf{C}ross-domain rec\textbf{O}mmendation vi\textbf{A} u\textbf{S}er in\textbf{T}erest alignment}, i.e. \textit{COAST}, which endeavors to improve cross-domain recommendation  with partial user overlap, as shown in Figure~\ref{photo}(b). Unlike previous studies, we extract enough features from the affluent content data (comments, tags, user/item profiles) to form an implicit data set to capture more feedback. Meanwhile, we modernize the previous approach of separately training representations into a unified cross-domain heterogeneous graph to assimilate the cross-domain similarity of users and items. Targeted at overlapping users' interest alignment across multiple domains, we gain in-depth insights from both user-user and user-item perspectives. Specifically, for user-user interest alignment, we believe that users' behaviors in different domains are driven by the same interest distribution, thus encouraging all views of the user to possess similar interest distributions over \textit{K} interest representations, as shown in Figure~\ref{photo}(a). This not only allows the model to distinguish users at the instance level, but also mitigates conflicting interests in views of the same user. For user-item interests alignment, we contend that interacted items are a observation of user interests, and all user views should exhibit consistent preferences for them. Particularly, benefiting from the rich semantics of gradients~\cite{gao2021gradient}, we employ gradient alignment to encourage higher-order projections across views to follow the same optimization path.

In this paper, we make the following contributions:
\begin{itemize}[leftmargin=12pt]
    \item To the best of our knowledge, we make significant efforts in cross-domain recommendation by considering cross-domain similarity and user interest alignment. Our framework performs dual knowledge transfer on basis of partial user overlap to improve recommendation performance.
    \item Instead of training entity representations separately, we construct a unified cross-domain heterogeneous graph, and correspondingly develop a novel message passing mechanism to capture the cross-domain similarity between entities.
    \item We resort to contrastive learning and gradient alignment to constrain user-user and user-item interest alignment, respectively, thereby enhancing the interest consistency across views.
    \item We compare \textit{COAST} to state-of-the-art algorithms for real-world recommendations, achieving significant improvements on all tasks. 
    We promise the code and data sets will be released for further comparison after acceptance~\footnote{https://github/anonymous/COAST}.
\end{itemize}

The rest of this paper is organized as follows. Section 2 briefly introduces related work, and then introduces the details of our proposed model. The experimental results and analysis are given
in Section 4. Finally, we summarize the paper in the fifth section.

\section{Related Work}
Our proposed framework stems from two research areas: cross-domain recommendation~\cite{zang2021survey} and contrastive learning~\cite{jaiswal2020survey}. We respectively summarize their main research paradigms, pros and cons, and close links with our research.
\subsection{Cross-domain Recommendation}
Cross-domain recommendation strives to explore data from multiple domains to simultaneously improve the recommendation performance of the model in all scenarios~\cite{khan2017cross}.

A rudimentary idea is to incorporate several constraints of cross-domain knowledge to decompose the user-item interaction matrices in both domains simultaneously~\cite{zhao2018low,he2018robust,wang2021cross}. This genre can be extended on a large number of matrix factorization-based single-domain recommendations~\cite{wu2022survey}, whereas its performance is inferior to deep learning approaches. Another paradigm is to customize a mapping function whose optimization objective is that the transformed cold-start user representation generalizes well in the target domain~\cite{zhu2021transfer,wang2021low}. The effiency of this paradigm depends on the reasonableness and representational power of the mapping function and whether enough overlapping entities are available for training, which limits the generalizability of the model. The third paradigm resorts to the popular knowledge graph technology~\cite{zhang2019heterogeneous}, which builds shared graphs to represent the relationships among users, items, and attributes, and learns entity representations through graph embeddings~\cite{li2020heterogeneous,cui2020herograph}. Despite the excellent extraction capability of graph structure, the high demands of computational resources make the scalability of these methods potentially limited.
Recently, algorithms utilizing overlapping user representations and combinations is trendy, and their standard practice is to learn entity representations from various domains, and then combine overlapping entity representations to enrich the knowledge of each domain~\cite{gao2019cross,zhu2021unified}. Apparently, the lack of cross-domain similarity and the rough combination way limit their recommendation performance.

Our approach falls within the last paradigm, but strives to conquer the proposed drawbacks. The closest algorithm to ours in this paradigm is GADTCDR~\cite{zhu2020graphical}, but they are fundamentally different.
First, at the data level, apart from explicit interactions, we attach exploration of content information. Second, at the algorithm level, we construct a unified cross-domain heterogeneous graph and user interest alignment for training, which enhances the generalization of the model. Finally, at the optimization level, we optimize in an end-to-end manner, avoiding the potential target inconsistency brought by two-stage training.

\subsection{Contrastive Learning}
Contrastive learning emphasizes learning common features between different views of an instance, with the intention of instance-level discrimination~\cite{liu2021self}. In contrast to supervised learning, it learns in a self-supervised manner.

Early contrastive learning architectures favored large batch sizes to aggregate enough negative examples, but the scalability of such methods was limited by GPU memory~\cite{chen2020simple}.
Aiming to improve on this,~\citet{wu2018unsupervised} applied a memory bank to store a large number of sample representations as negative examples, thus avoiding the common out-of-memory. Despite the approximate performance, a potential pitfall of this approach is that representation updates in the memory bank can be computationally expensive as it becomes outdated quickly within a few iterations. Consequently,~\citet{he2020momentum} further improved the form of the static repository, using a momentum encoder to generate a dictionary as a queue for encoding keys, the current mini-batch is enqueued, and the oldest mini-batch is dequeued. This approach eliminates the need to use two separate models for feature extraction, and dynamic queues avoid excessive memory consumption. All of the above architectures place insight into using specific metrics to measure sample similarity, i.e., encouraging different views of the same entity to be closer in the projected space and vice versa~\cite{wu2021self}. Recently,~\citet{caron2020unsupervised} abandoned the traditional comparison of positive and negative examples, and launched a new exploration of contrastive learning from the perspective of clustering.

Inspired by contrastive learning, we intend to discriminate user representations at the instance level. Particularly, following the idea of clustering, we encourage different views of the same user to aggregate into the same interest center, thereby generating better user interest representations.

\section{Proposed Method}
In this section, we introduce the proposed \textit{COAST} framework. Specifically, we first elaborate the definition of the general CDR problem, then outline our framework, and finally detail the sub-modules and optimization methods.


\subsection{Problem Formulation}
This work considers a general CDR scenario with two domains $\mathcal{S}$ (source) and $\mathcal{T}$ (target), where the former contains rich and informative interactions and the latter is relatively sparse. Suppose source domain $\mathcal{D_{S}=(U_{S}, V_{S}, E_{S}, X_{S})}$, target domain $\mathcal{D_{T}=(U_{T}, V_{T},}$ $\mathcal{E_{T},X_{T})}$, where $\mathcal{U, V, E, X}$ are user set, item set and edge set, attribute set in each domain, respectively. 
In particular, the user sets $\mathcal{U_{S}}$ and $\mathcal{U_{T}}$ contain an overlapping user subset $\mathcal{U}_{o}$. Then, the user set can be redefined as $\mathcal{U_{S}}=\{\mathcal{U}_{s},\mathcal{U}_{o}\}$, $\mathcal{U_{T}}=\{\mathcal{U}_{t},\mathcal{U}_{o}\}$, where $\mathcal{U}_{s}$ and $\mathcal{U}_{t}$ are non-overlapping/distinct user sets in the two domains. For simplicity of exposition, we further introduce two binary matrices to store user-item interactions, namely $\mathcal{A_S}=\{0,1\}^\mathcal{{|U_S|\times|V_S|}}$, $\mathcal{A_T}=\{0 ,1\}^\mathcal{{|U_T|\times|V_T|}}$, where element $A_{ij}$ in each domain denotes whether the user $u_{i} \in \mathcal{U}$ and item $v_{j} \in \mathcal{V}$ have an interaction in the edge set $\mathcal{E}$. The definition of dual cross-domain recommendation is as follows,

\textit{Given the observed interaction and content of $\mathcal{S}$ and $\mathcal{T}$, dual CDR aims to leverage knowledge transfer from overlapping users to improve recommendation performance in both domains.
Formally, given $\mathcal{A_S}$, $\mathcal{A_T}$, $\mathcal{X_{S}}$, $\mathcal{X_{T}}$ , we expect to recommend $v_{i}\in \mathcal{V_{S}}$, $v_{j}\in \mathcal{V_{T}}$ respectively in domains $\mathcal{S}$ and $\mathcal{T}$.}

Important mathematical notes can be found in Appendix~\ref{A}.

\subsection{Overview of \textit{COAST} Framework}
\begin{figure}[!ht]
\centering
\includegraphics[width=\linewidth]{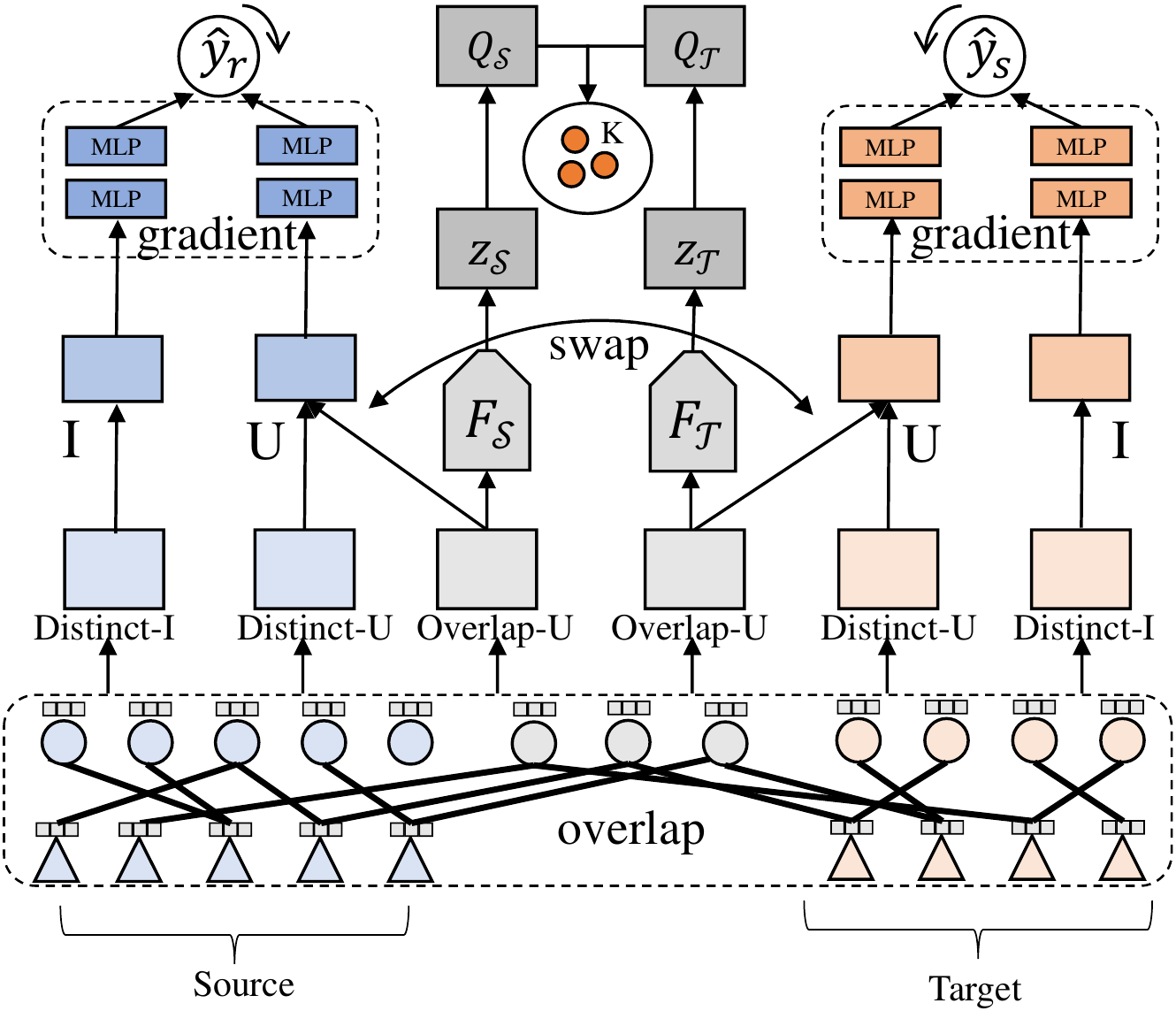}
\centering
\caption{Overall framework of \textit{COAST}.}
\label{frame}
\end{figure}
In this section, we outline the proposed cross-domain recommendation framework \textit{COAST}, whose architecture is shown in Figure~\ref{frame}. First, we construct a unified cross-domain heterogeneous graph, and improve the message passing mechanism of graph convolutional network to capture the cross-domain similarity of users and items. Then, for each overlap user, we utilize contrastive learning and gradient alignment from both user-user and user-item perspectives to ensure the alignment of user interests. Finally, following previous studies, we adopt a negative sampling mechanism to calculate the supervision loss of the two domains, which is jointly optimized with the above two losses for alignment.

\subsection{Cross-domain Graph Convolution}
We argue that previous separately trained representations can only capture single-domain information; therefore we construct a unified cross-domain heterogeneous graph and a novel message passing mechanism to capture cross-domain similarity.
\subsubsection{Construction}
We determine nodes and edges in the heterogeneous graph $\mathcal{G}$ on basis of $\mathcal{A_S}$ and $\mathcal{A_T}$. Note that for items from both domains, we treat them as nodes of the same type, the difference being the type of edges users interact with them. For the initial embeddings of nodes, we generate them in the following data preprocessing manner. Specifically, for common numerical attributes and category attributes, we perform normalization and one-hot encoding respectively. For text attributes (tags, comments, profiles, etc.), we first aggregate the text associated with the entity into a large document, which is then converted into semantic vectors using doc2vec technique~\cite{dai2015document}. Note that we perform joint encoding on users of both domains. Finally, we get the initial embedding for each user and item, i.e., $e^{u}\in \mathcal{H_{U}}$, $e^{v}_\mathcal{S}\in \mathcal{H_{S}}$, $e^{v}_\mathcal{T}\in \mathcal{H_{T}}$. Formally,
\begin{equation}
	e^{u} = \begin{cases}
	      e^{u}_{\mathcal{S}} , & if \ u \in \mathcal{U}_s \\
	      e^{u}_{\mathcal{T}}, & if \ u \in \mathcal{U}_t \\
            e^{u}_{\mathcal{S}} \otimes e^{u}_{\mathcal{T}}, & if \ u \in \mathcal{U}_o \\
		   \end{cases},
\end{equation}
where $\otimes$ is max pooling operation. Overlapping users have behaviors in both domains, so we aggregate their representations in both domains. Without loss of generality, we adopt max pooling here. We experimented with operations such as sum and averaging, and found no significant improvement. 
\subsubsection{Propagation}\label{inters}
To capture the high-order cross-domain similarity of users and items, we improve upon the message passing mechanism of graph convolution networks~\cite{wang2019neural}. Formally,
\begin{equation}
m_{u \leftarrow v}=\frac{1}{\sqrt{|\mathrm{N}_{u}||\mathrm{N}_{v}^{\mathcal{S}}||\mathrm{N}_{v}^{\mathcal{T}}|}}(W_{1} e^{u} + W_{2}(e^{v}_{\mathcal{S}} \odot e^{u}) +  W_{3}(e^{v}_{\mathcal{T}} \odot e^{u})),
\end{equation}
where $\mathrm{N}$ represents set of 1-hop neighbors, $W$ is a trainable parameter, and $\odot$ denotes the element-wise product. We add cross-domain user-item interactions to the message passing mechanism of graph convolution operation, expecting to capture historical interaction information. This approach not only enriches the embedding representation, but also enhances the capture of cross-domain collaborative signals. Formally, the user embedding propagation is,
\begin{equation}
e^{u^{(l+1)}}=\operatorname{LeakyReLU}(m_{u \leftarrow u}^{(l)}+\sum_{v \in \mathrm{N}_{u}} m_{u \leftarrow v}^{(l)}),
\end{equation}
where $l$ represents the $l$-th GNN layer. We also support stacking of GNN layers to perceive higher-order similarities. Formally,

\begin{equation}
\begin{split}
E^{(l)}=\sigma((L+I) E^{(l-1)} W_1^{(l)} &+ LE^{(l-1)} \odot E^{(l-1)} W_2^{(l)} \\
&+ LE^{(l-1)} \odot E^{(l-1)} W_3^{(l)}),
\end{split}
\end{equation}
where $\sigma$ is activation function $Relu$, $E$ is the representations for users and items, $I$ denotes an identity matrix. $L$ represents the Laplacian matrix for the graph. Formally,

\begin{equation}
L= \mathrm{D}^{-\frac{1}{2}}A \mathrm{D}^{-\frac{1}{2}} \text { and } A=\left[\begin{array}{cc}
\mathbf{0} & R \\
R^{\top} & \mathbf{0}
\end{array}\right],
\end{equation}
where $\mathrm{D}$ is the diagonal degree matrix, $A$ is the adjacency matrix, $R$ is the user-item interaction matrix and $\mathbf{0}$ is all zero matrix. We concat the user and item representations of each layer, i.e., $e^{u}=e^{u^{(0)}} \oplus \cdots \oplus e^{u^{(l)}}, e^{v}_{\mathcal{S}}=e^{v^{(0)}}_{\mathcal{S}} \oplus \cdots \oplus e^{v^{(l)}}_{\mathcal{S}}, e^{v}_{\mathcal{T}}=e^{v^{(0)}}_{\mathcal{T}} \oplus \cdots \oplus e^{v^{(l)}}_{\mathcal{T}}$.

Our approach has several advantages. On the one hand, we form a unified graph structure for user-item interactions in different domains, which is intuitive and easy to capture cross-domain similarity. On the other hand, we generalize the message passing mechanism to cross-domain scenarios, enhancing the practicality of traditional graph convolution operators.

\subsection{User Interest Alignment}
Previous studies applied plain representation aggregation to transfer knowledge of both domains; however we argue that this approach ignores the alignment of user interests.
Consequently, we align user interests from user-user and user-item perspectives to constrain user representation.
\subsubsection{User-User Alignment}~\label{uu}
\begin{figure}[!ht]
\centering
\includegraphics[width=\linewidth]{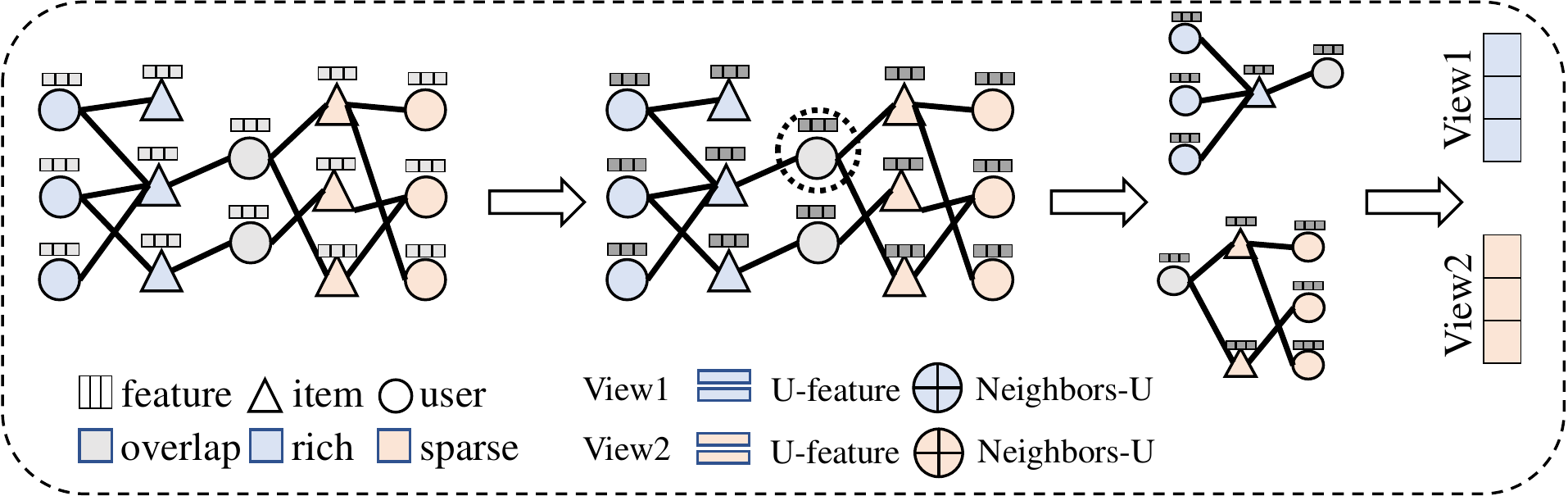}
\centering
\caption{User-User interest alignment.}
\label{contrast}
\end{figure}
To discriminate users at the instance level, we separately aggregate users’ second-order neighbors’ representation in different domains to obtain corresponding contrastive views, as shown in Figure~\ref{contrast}. The motivation behind it is that the user's context can enhance the user's interest representation in this domain, which is widely used in single-domain graph recommendation~\cite{liu2021contextualized}. Formally, for $ u \in \mathcal{U}_o$
\begin{equation}
    e^{u}_{\mathcal{S}} = \sum_{i \in {\mathrm{N}_{u,2}^{\mathcal{S}}} }\alpha_{i} e^{u_{i}},  \qquad
    e^{u}_{\mathcal{T}} = \sum_{i \in {\mathrm{N}_{u,2}^{\mathcal{T}}}} \alpha_{i} e^{u_{i}},
\end{equation}
where $\mathrm{N}_{u,2}$ is the 2-hop neighbors of $u$ and  $\alpha_{i} = \frac{\mathrm{exp}(s(u_{i},u))}{\sum_{j \in \mathrm{N}_{u,2}} \mathrm{exp}(s(u_{j},u))}$. $s(\cdot)$ represents the scoring function, and without loss of generality, we use the dot product.

Then we feed $e^{u}_{\mathcal{S}}$ and $e^{u}_{\mathcal{T}}$ into the feature extractors $F_\mathcal{S}$ and $F_\mathcal{T}$ respectively, and get their representations $z_\mathcal{S}$, $z_\mathcal{T}$. We assume that overlapping users have a total of \textit{K} interests, i.e., $\{c_{1},\cdots, c_{K}\}$. According to our assumption, the distribution of interests among different views of the same user should be consistent. Formally,

\begin{equation}
\begin{split}
\ell(z_{\mathcal{T}}, q_{\mathcal{S}}) &= -\sum_k q_{\mathcal{S}}^{(k)} \log p_{\mathcal{T}}^{(k)} \\
p_{\mathcal{T}}^{(k)} &= \frac{\exp (\frac{1}{\tau} z_{\mathcal{T}}^{\top} c_k)}{\sum_{k^{\prime}} \exp (\frac{1}{\tau} z_{\mathcal{T}}^{\top} c_{k^{\prime}})}
\end{split},
\end{equation}
where $q$ is the higher-order projection through the Q extractor and 
$\tau$ is a temperature parameter. In other words, we encourage the contrastive views of $u$ to posses the same clustering results over interest distribution.
The user-user alignment loss is as follows,
\begin{equation}
\mathcal{L}_{\mathcal{U},\mathcal{U}}=\ell(z_{\mathcal{T}}, q_{\mathcal{S}})+\ell(z_{\mathcal{S}}, q_{\mathcal{T}}),
\end{equation}
Moreover, we follow the same solution in swav~\cite{caron2020unsupervised},
which restricts the transportation of tensors in the mini-batch to ensure that the model is memory efficient.

\subsubsection{User-Item Alignment}
\begin{figure}[!ht]
\centering
\includegraphics[width=\linewidth]{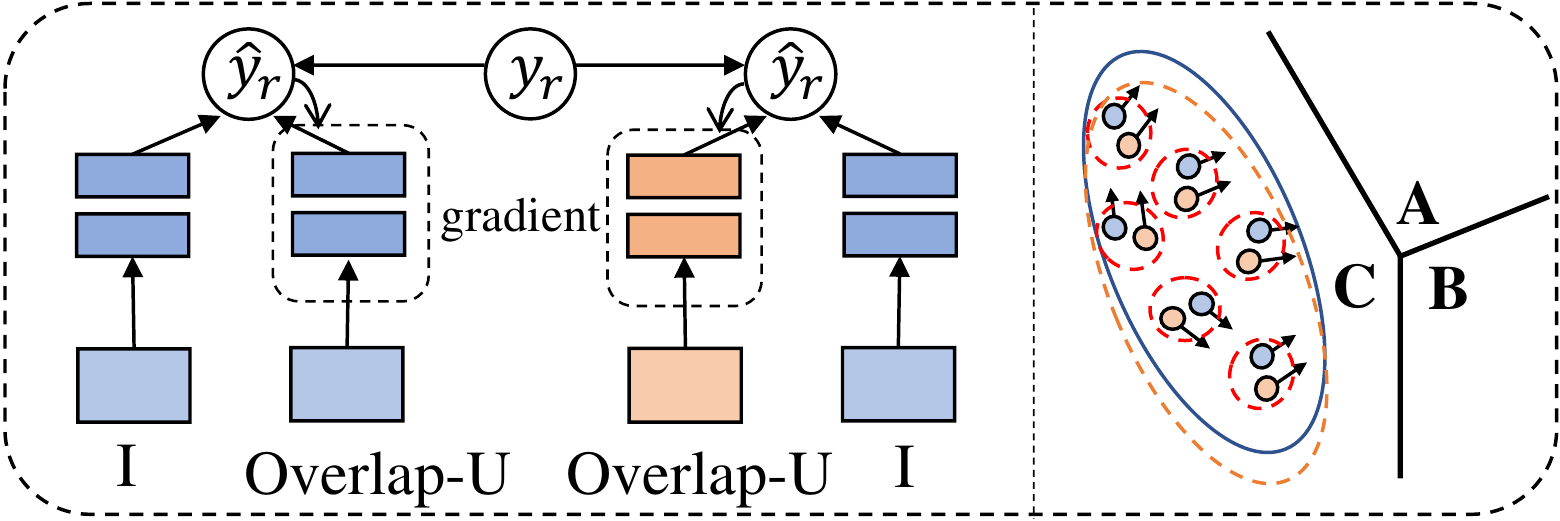}
\centering
\caption{User-Item interest alignment.}
\label{gradient}
\end{figure}

To ensure consistent user interest in items, we encourage different views of $u$ to be closer to the interacted item representation, as shown in Figure~\ref{gradient}.
A straightforward motivation of this insight is that both user views and interacted items can represent the user's real interests; therefore they should be close in the projected space, even if the views and items are in different domains. Consequently, benefit from the rich semantics of gradients~\cite{gao2021gradient}, we introduce gradient alignment to induce different views to follow the same optimization path for interacted items. Formally, we define $g_{\mathcal{S}}$ and $g_\mathcal{T}$ to represent the expected gradients on the user's source and target views.
\begin{equation}
g_{\mathcal{S}} = \underset{(u, v) \sim(\mathcal{U}_{o}, \mathcal{V_{S}})}{\mathrm{E}}[\nabla_{\theta_{f_{s}^{u}}} \ell_{ce}(F^{u}_{s}(e^{u}) \cdot (F^{v}_{s}(e^{v}))^{\top}, y_{u,v})],
\end{equation}
where $F^{u}_{s}$, $F^{v}_{s}$ are tower structures for extracting the representations of users and items in the source domain, both composed of Multi-Layer Perceptrons (MLPs).
\begin{equation}
g_{\mathcal{T}} = \underset{(u, v) \sim(\mathcal{U}_{o}, \mathcal{V_{T}})}{\mathrm{E}}[\nabla_{\theta_{f_{t}^{u}}} \ell_{ce}(F^{u}_{t}(e^{u}) \cdot (F^{v}_{t}(e^{v}))^{\top}, y_{u,v})],
\end{equation}
We aim to minimize discrepancy between $g_{\mathcal{S}}$ and $g_{\mathcal{T}}$. Without loss of generality, we use cosine similarity as the discrepancy measure.
\begin{equation}
\mathcal{L}_{U,I}=1-\frac{g_{\mathcal{S}}^{\top} \cdot g_{\mathcal{T}}}{\|g_{\mathcal{S}}\|_2\|g_{\mathcal{T}}\|_2},
\end{equation}
where $|| \cdot ||_2$ represents the 2-Norm.

Overall, we constrain user representations from a more fine-grained perspective, i.e., user interest alignment. On the one hand, this approach acts as a regularizer to prevent overfitting of user representations. On the other hand, contrastive learning utilizes unsupervised information and gradient alignment utilizes semantic information, both of which further enrich the transfer of cross-domain knowledge.
\subsection{Model Optimization}
In this section, we first elaborate the supervised prediction of \textit{COAST}, and then illustrate the joint optimization process.
\subsubsection{Supervised Estimation}
Similar to the previous work~\cite{li2020ddtcdr}, we adopt a dual-tower structure to capture high-order representations of users and items, where the tower structure is composed of MLPs. The structure of MLPs uses $[D, 2D, 4D, 8D, 4D, 2D, D]$, which has been shown to be effective in feature extraction~\cite{zhu2020graphical}.
\begin{equation}
\begin{split}
\hat{y_s} = \frac{F^{u}_{s}(e^{u}_{\mathcal{S}}) \cdot (F^{v}_{s}(e^{v}_{\mathcal{S}}))^{\top}}{||F^{u}_{s}||||F^{v}_{s}||} + \lambda_{1}(||e^{u}||+||e^{v}_{\mathcal{S}}||)  \\
\hat{y_t} = \frac{F^{u}_{t}(e^{u}_{\mathcal{T}}) \cdot (F^{v}_{t}(e^{v}_{\mathcal{T}}))^{\top}}{||F^{u}_{t}||||F^{v}_{t}||} + \lambda_{1}(||e^{u}||+||e^{v}_{\mathcal{T}}||) 
\end{split},
\end{equation}
where $||e^{u}||$ is the embedding regularizer. To avoid our model overfitting $Y^{+}$ (ground truth), we randomly select a certain number of unobserved user-item interactions as negative instances, denoted $Y^{-}$, $y=\{Y^{+},Y^{-}\}$. This negative sampling-based training strategy has been widely used in existing algorithms~\cite{zhou2019deep}. Formally, we optimize using binary cross-entropy,
\begin{equation}
\ell(y, \hat{y})=y \log \hat{y}+(1-y) \log (1-\hat{y}),
\end{equation}
The supervised loss is optimized in both domains simultaneously,
\begin{equation}
\mathcal{L}_{s} = \ell(y_{s}, \hat{y_{s}}) +  \ell(y_{t}, \hat{y_{t}})
\end{equation}

\subsubsection{Total Loss}
Loss functions for each part are added together for joint optimization. The overall loss function is
\begin{equation}
\mathcal{L}=\mathcal{L}_{s}+\lambda_{2} (\mathcal{L}_{U,U}+ \mathcal{L}_{U,I}),
\end{equation}
where $\lambda_2$ is the weight of the two interest alignment constraints. 

Overall, we propose an end-to-end solution for dual cross-domain recommendation, which can improve the recommendation performance of both domains while ensuring the alignment of overlapping user interests.
The overall optimization process of the algorithm is shown in Algorithm~\ref{alg1} in Appendix~\ref{B}.

\section{Experiments}
To demonstrate the state-of-the-art and robustness of our model, we conduct extensive experiments to answer the following questions: 
\begin{itemize}[leftmargin=12pt]
    \item RQ1: How does \textit{COAST} perform on common metrics compared to state-of-the-art algorithms?
    \item RQ2: How do overlapping user ratios and sub-modules affect model performance?
    \item RQ3: What impact do several key parameters have on model performance?
\end{itemize}

\subsection{Experimental Settings}
In this section, we present the statistics of the data sets, necessary parameter settings for the model, and state-of-the-art algorithms for comparison.
\subsubsection{Data Sets}
We conduct extensive experiments using large-scale anonymized data sets obtained from Douban and a well-known industrial platform. They both allow users to rate and review a range of items from different domains, each of which represents the user's interests. On that account, the combination of explicit user feedback and implicit domain knowledge is unique and valuable for cross-domain recommendation.
\begin{itemize}[leftmargin=12pt]
    \item \textbf{Douban data set.} We choose a subset containing the three largest domains, including books, movies, and music. They are linked together by a shared user ID that identifies the same user. Correspondingly, we construct three cross-domain recommendation tasks: \textit{movie-book}, \textit{movie-music}, and \textit{book-music}.
    \item \textbf{Industrial data set.} This platform has two scenarios, mall and community, which are connected by a shared user id. Consequently, we constructed a task \textit{mall-community}, expecting to improve the recommendation performance in both domains.
\end{itemize}

\begin{table}[!h]
\centering
\caption{Statistics of data sets.}
\label{stat}
\setlength{\tabcolsep}{1mm}{
\begin{tabular}{c|c|c|c|c|c} 
\hline
data sets     & \multicolumn{3}{c|}{Douban} & 
\multicolumn{2}{c}{Industrial Platform}  \\ 
\hline
Domains      & Movie   & Music  & Book     & Mall & Community            \\ 
\hline
Users        & 2,712    & 1,672   & 2,110     & 35,233     &  29,355                    \\ 
\hline
Items        & 34,893   & 5,567   & 6,777     &  1,749   
&  2,452                    \\ 
\hline
Interactions & 1,278,401 & 69,709  & 96,041    &    319,795  &   175,802                   \\ 
\hline
Density      & 1.35\%  & 0.75\% & 0.67\%   &  0.52\%    &  0.24\%                    \\
\hline
\end{tabular}}
\setlength{\tabcolsep}{1mm}{
\begin{tabular}{c|c|c|c|c} 
\hline
\multicolumn{2}{c|}{\textbf{Tasks }}     & \textbf{Richer} & \textbf{Sparser} & \textbf{Overlap}  \\ 
\hline
\multirow{3}{*}{Douban} & \textbf{Task1} & Movie           & Book             & 2,106              \\ 
\cline{2-5}
                        & \textbf{Task2} & Movie           & Music            & 1,666              \\ 
\cline{2-5}
                        & \textbf{Task3} & Book            & Music            & 1,566              \\ 
\hline
Industrial Platform                  & \textbf{Task4} & Mall            & Community        & 3,146              \\
\hline
\end{tabular}}
\end{table}
Statistics on the two data sets can be found in Table~\ref{stat}.
For both data sets, the user's content features are aggregated by user comments, user tags, and user profiles, and the item's content features are composed of its profile and the comments below it. Note that each user may interact with items from different domains, but each item belongs to only one domain. To improve data quality, we filter all data sets to keep users and items with at least 5 interactions~\cite{zhu2020graphical}. We normalize the scoring range from 0 to 1.

\subsubsection{Parameter Settings}
Our framework is implemented using Pytorch. Except for the necessary concat operation, the embedding size is 64. We adopt Kaiming method~\cite{he2015delving} for parameter initialization. For gradient descent, we take Adam~\cite{kingma2014adam} with the initial learning rate 5e-4 for model optimization. In our proposed model, we set batch size to 4096 and the training maximum epoch to 100. We initialize the user's interest \textit{K} to 256, set the regularization weight $\lambda_1$ and alignment weight $\lambda_2$ to 1e-2 and 1e-3, respectively. 

Similar to previous work~\cite{deng2020personalized}, we adopt a \textit{leave-one-out} approach to evaluate model performance. Specifically, for each user in the test set, we randomly sample 99 items that the user has not interacted with as negative examples, and calculate the ground truth hit rate and ranking position. The results of model and baselines are evaluated by Hit Ratio (Hit) and Normalized Discounted Cumulative Gain (NDCG) values, where HR measures whether the test item is ranked on the Top-N list while NDCG measures the specific ranking quality that assigns high scores to hits at top position ranks~\cite{he2017neural}. Note that this paper is evaluated with @10 unless otherwise specified.
\subsubsection{Baselines}
To verify the effectiveness of cross-domain recommendation and the superiority of our model, we choose the classic single-domain recommendation algorithms and cross-domain recommendation approaches for comparison.
\begin{itemize}[leftmargin=12pt]
	\item \textbf{NMF~\cite{he2017neural}:} NMF aims to find a reasonable user-item interaction function for recommendation by combining the linearity of MF and the nonlinearity of MLP.
	\item \textbf{LightGCN~\cite{he2020lightgcn}:} LightGCN only obtains node embeddings by neighborhood aggregation because it believes that feature transformation and nonlinear activation have little effect on collaborative filtering, and even damage recommendation performance.
	\item \textbf{MVDNN~\cite{elkahky2015multi}:} MVDNN maps users and items from multiple domains into a common latent space, and optimizes by maximizing the similarity between users and their preferred items.
	\item \textbf{DTCDR~\cite{zhu2019dtcdr}:} DTCDR extends NMF to cross-domain recommendation, leveraging the textual and rating representations of overlapping users from both domains for knowledge transfer.
	\item \textbf{DDTCDR~\cite{li2020ddtcdr}:} DDTCDR seeks to learn a latent orthogonal mapping function between domains to obtain cold-start user representations in other domains.
      \item \textbf{DML~\cite{li2021dual}:} DML further extends DDTCTR based on dual metric learning, which exploits multiple orthogonal mapping functions to explore the transfer of cold-start user representations.
     \item \textbf{GADTCDR~\cite{zhu2020graphical}:} GADTCDR adds user-user and item-item edges to heterogeneous graphs based on content similarity to improve representation capabilities.
     \item \textbf{CDRIB~\cite{cao2022cross}:} CDRIB uses the information bottleneck principle to  debias recommendations in two domains.
\end{itemize}
Please note that NMF and LightGCN are single-domain recommendation algorithms, and experiments are performed on the two domains separately. The others are cross-domain recommendation algorithms, where DDTCDR and DML are mapping-based methods, while MVDNN, DTCDR, GADTCDR, and CDRIB are representation-combination-based approaches.
To be fair, we tune the hyper-parameters of each model to achieve the best results.
\subsection{Comparison with Baselines (RQ1)}


\begin{table*}
\centering
\caption{Performance comparison for cross-domain recommendation.}\label{com}
\resizebox{\textwidth}{!}{
\begin{threeparttable}
\begin{tabular}{c|c|c|c|c||c|c|c|c||c|c|c|c||c|c|c|c} 
\toprule
\multirow{3}{*}{Algorithm} & \multicolumn{4}{c||}{Task1}                                           & \multicolumn{4}{c||}{Task2}                                           & \multicolumn{4}{c||}{Task3}                                            & \multicolumn{4}{c}{Task4}                                              \\ 
\cline{2-17}
                           & \multicolumn{2}{c|}{Movie}        & \multicolumn{2}{c||}{Book}        & \multicolumn{2}{c|}{Movie}        & \multicolumn{2}{c||}{Music}       & \multicolumn{2}{c|}{Book}         & \multicolumn{2}{c||}{Music}        & \multicolumn{2}{c|}{Mall}         & \multicolumn{2}{c}{Community}      \\ 
\cline{2-17}
                           & Hit             & NDCG            & Hit             & NDCG            & Hit             & NDCG            & Hit             & NDCG            & Hit             & NDCG            & Hit             & NDCG             & Hit             & NDCG            & Hit             & NDCG             \\ 
\hline
NMF                        & 0.5445          & 0.3154          & 0.3916          & 0.2224          & 0.5445          & 0.3154          & 0.3959          & 0.2206          & 0.3916          & 0.2224          & 0.3959          & 0.2206           & 0.5850          & 0.3265          & 0.3793          & 0.2048           \\ 
\hline
LightGCN                   & 0.6174          & 0.3492          & 0.3805          & 0.2226          & 0.6174          & 0.3492          & 0.3528          & 0.2023          & 0.3805          & 0.2226          & 0.3528          & 0.2023           & 0.5848          & 0.2933          & 0.5119          & 0.2490           \\ 
\hline
MVDNN                      & 0.6382          & 0.3689          & 0.4654          & 0.2575          & 0.6414          & 0.3641          & 0.3965          & 0.2238          & 0.5104          & 0.2947          & 0.3923          & 0.2390           & 0.5963          & 0.3002          & 0.5211          & 0.2507           \\ 
\hline
DTCDR                      & 0.6420          & 0.3794          & 0.4302          & 0.2394          & 0.6197          & 0.4278$\dagger$ & 0.3593          & 0.2211          & 0.5108$\dagger$ & 0.3263$\dagger$ & 0.2848          & 0.2017           & 0.5580          & 0.3109          & 0.3632          & 0.2643           \\ 
\hline
DDTCDR                     & 0.5937          & 0.3558          & 0.4436          & 0.2511          & 0.5921          & 0.3722          & 0.3467          & 0.2189          & 0.4540          & 0.2666          & 0.3086          & 0.2042           & 0.5135          & 0.2884          & 0.3729          & 0.1886           \\ 
\hline
DML                        & 0.6060          & 0.3638          & 0.4662          & 0.2662          & 0.6093          & 0.4059          & 0.3821          & 0.2287          & 0.4521~         & 0.2616~         & 0.4253$\dagger$ & 0.2548$\dagger$~ & 0.5491          & 0.3181          & 0.4283          & 0.2124           \\ 
\hline
GADTCDR                    & 0.6817$\dagger$ & 0.4205$\dagger$ & 0.4882$\dagger$ & 0.3026$\dagger$ & 0.6818$\dagger$ & 0.4276          & 0.4383$\dagger$ & 0.2498$\dagger$ & 0.4492          & 0.2761          & 0.3571          & 0.1933           & 0.6654$\dagger$ & 0.4055$\dagger$ & 0.5173$\dagger$ & 0.2907$\dagger$  \\ 
\hline
CDRIB                      & 0.6114          & 0.3301          & 0.4630          & 0.2772          & 0.6411          & 0.3578          & 0.4103          & 0.2272          & 0.5021          & 0.2654          & 0.2866          & 0.2038           & 0.5744          & 0.3007          & 0.4802          & 0.2814           \\ 
\hline
COAST                      & 0.6905          & 0.4271          & 0.5052          & 0.3174          & 0.6938          & 0.4292          & 0.4497          & 0.2515          & 0.5138          & 0.3293          & 0.4688          & 0.2712           &   0.6769              &  0.4073               &    0.5503             &    0.3195              \\ 
\hline
Improvement                & 1.2909\%        & 1.5696\%        & 3.4821\%        & 4.8909\%        & 1.7600\%        & 0.3273\%        & 2.6001\%        & 0.6805\%        & 0.5873\%        & 0.9194\%        & 10.2280\%       & 6.4364\%         & 1.7283\%                &       0.4439\%          &  6.3793\%               &   9.9071\%          \\     
\bottomrule
\end{tabular}
\begin{tablenotes}
 \item[] $\dagger$ means the strongest baseline's performance.
\end{tablenotes}
\end{threeparttable}
}
\end{table*}

The results of all algorithms on the four tasks are shown in Table~\ref{com}, with the last row representing the improvement of our model over the best baseline for that task.
To summarize, benefiting from perception of cross-domain similarity and user interest alignment, \textit{COAST} achieved 0.32\%-10.22\% improvement compared to the best performance on different tasks. 

These experiments reflect some interesting findings: (1) Cross-domain algorithms outperform single-domain algorithms in most tasks, demonstrating the importance of knowledge transfer in cross-domain recommendation. Underperforming cross-domain baselines, especially those based on mapping genres, over-rely on overlapping user ratios such as DDTCDR, DML. (2) Algorithms incorporating implicit features outperform models using only explicit interactions, indicating the importance of capturing content similarity. (3) The representation-combination-based models outperform the mapping-based approaches, proving that a custom simple mapping function cannot reflect the complex transformation of user representations across domains.
(3) The improvement of the target domain is greater than that of the source domain. On the one hand, the source domain can provide more information, and on the other hand, the improvement of the recommendation capability of the source domain leads to a further promotion in the upper bound of the recommendation performance of the target domain.
(4) Furthermore, we observe that the proposed model improves the \textit{movie-book} task larger than the \textit{movie-music} task. The possible reasons are differences in data set size and the number of overlapping users, which determine the richness of knowledge and the caliber of transfer. We plan to leave this as a topic for future research.

\subsection{Robust Testing (RQ2, RQ3)}
We perform overlap ratio tests, ablation experiments, and hyper-parameter tests to verify the robustness of our model.
\subsubsection{Length N}
\begin{figure}
\centering
\subfigure[Hit of Douban-movie.]{
\begin{minipage}[t]{0.45\linewidth}
\centering
\includegraphics[width=\linewidth,height=0.75\linewidth]{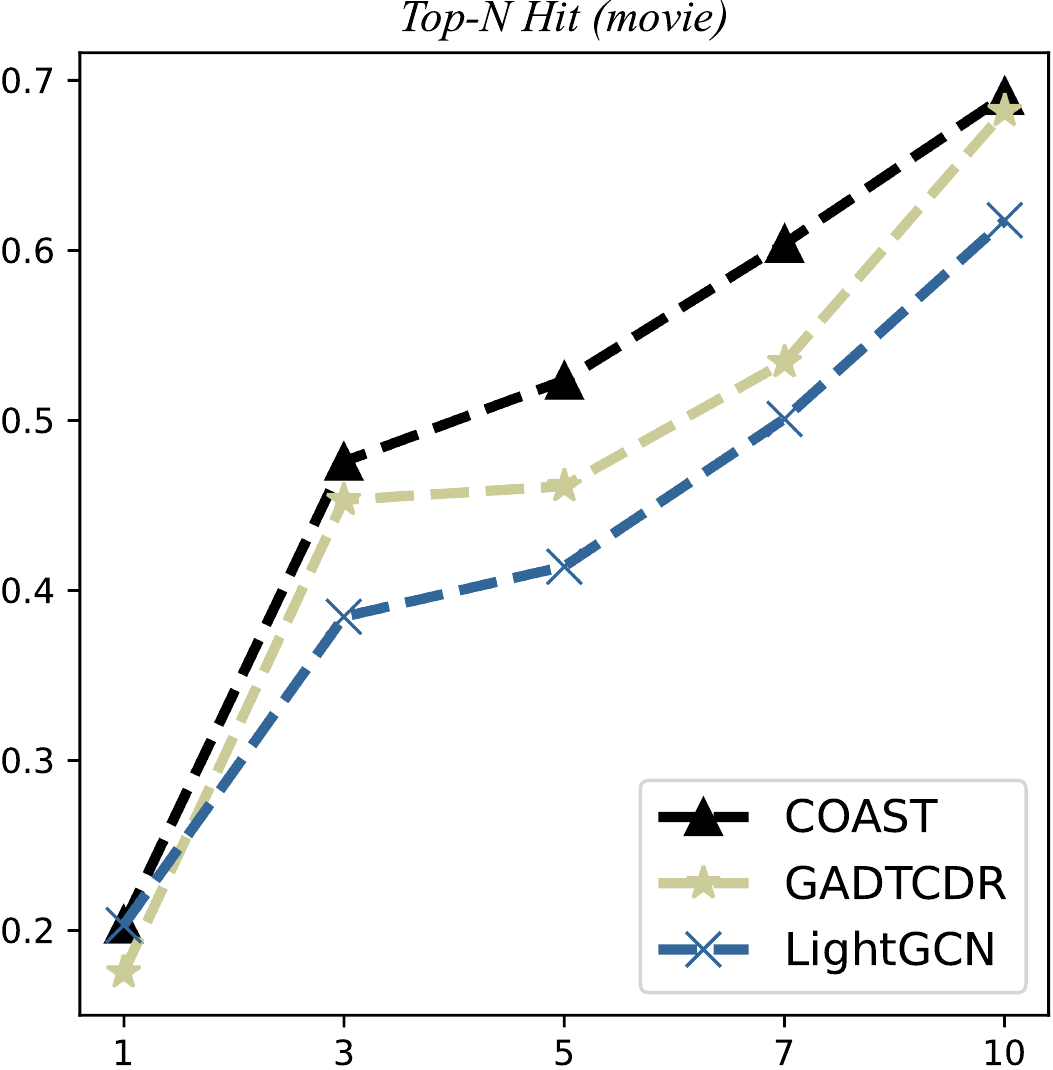}
\end{minipage}%
}%
\subfigure[NDCG of Douban-movie.]{
\begin{minipage}[t]{0.45\linewidth}
\centering
\includegraphics[width=\linewidth,height=0.75\linewidth]{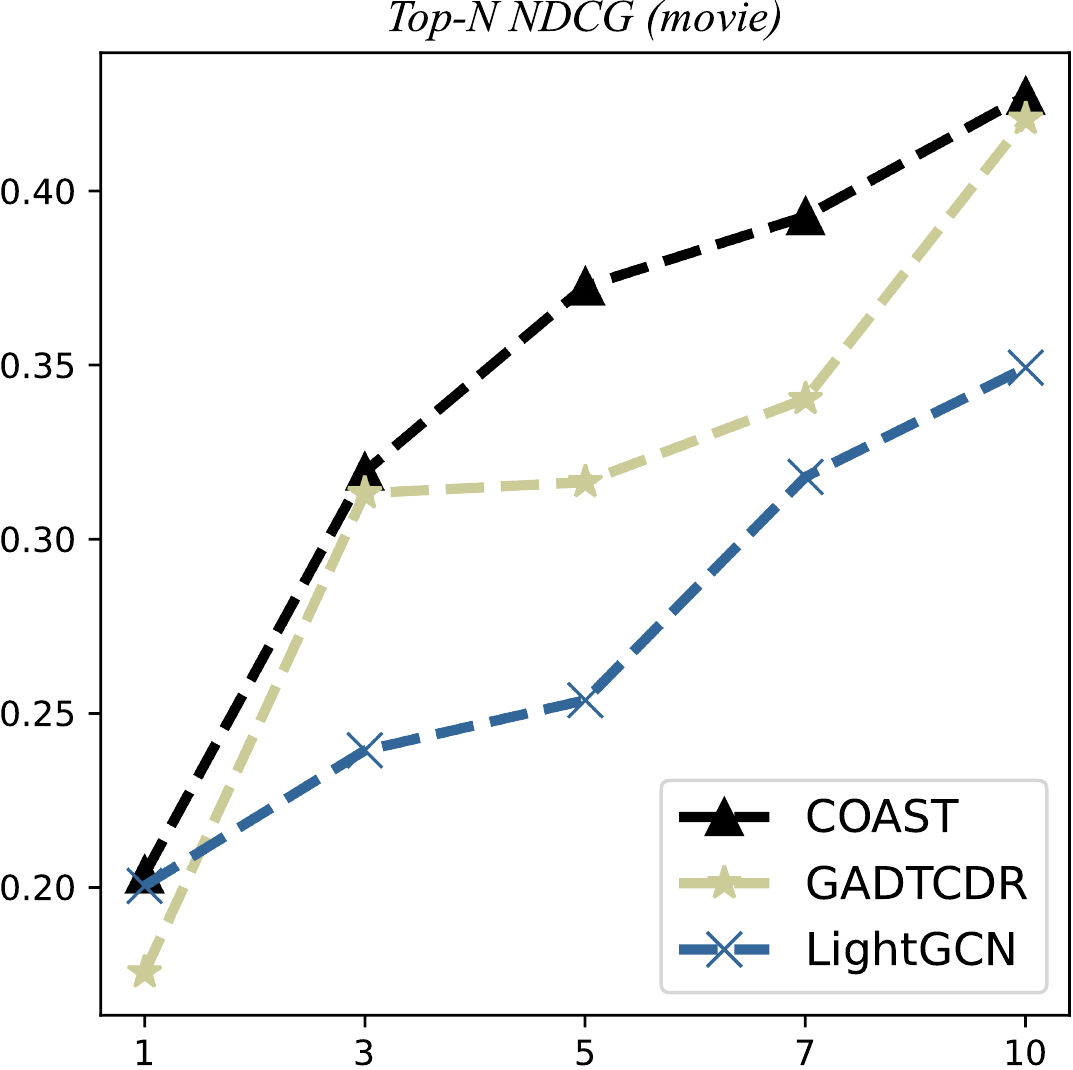}
\end{minipage}%
}%
\vskip\baselineskip
\subfigure[Hit of Douban-book.]{
\begin{minipage}[t]{0.45\linewidth}
\centering
\includegraphics[width=\linewidth,height=0.75\linewidth]{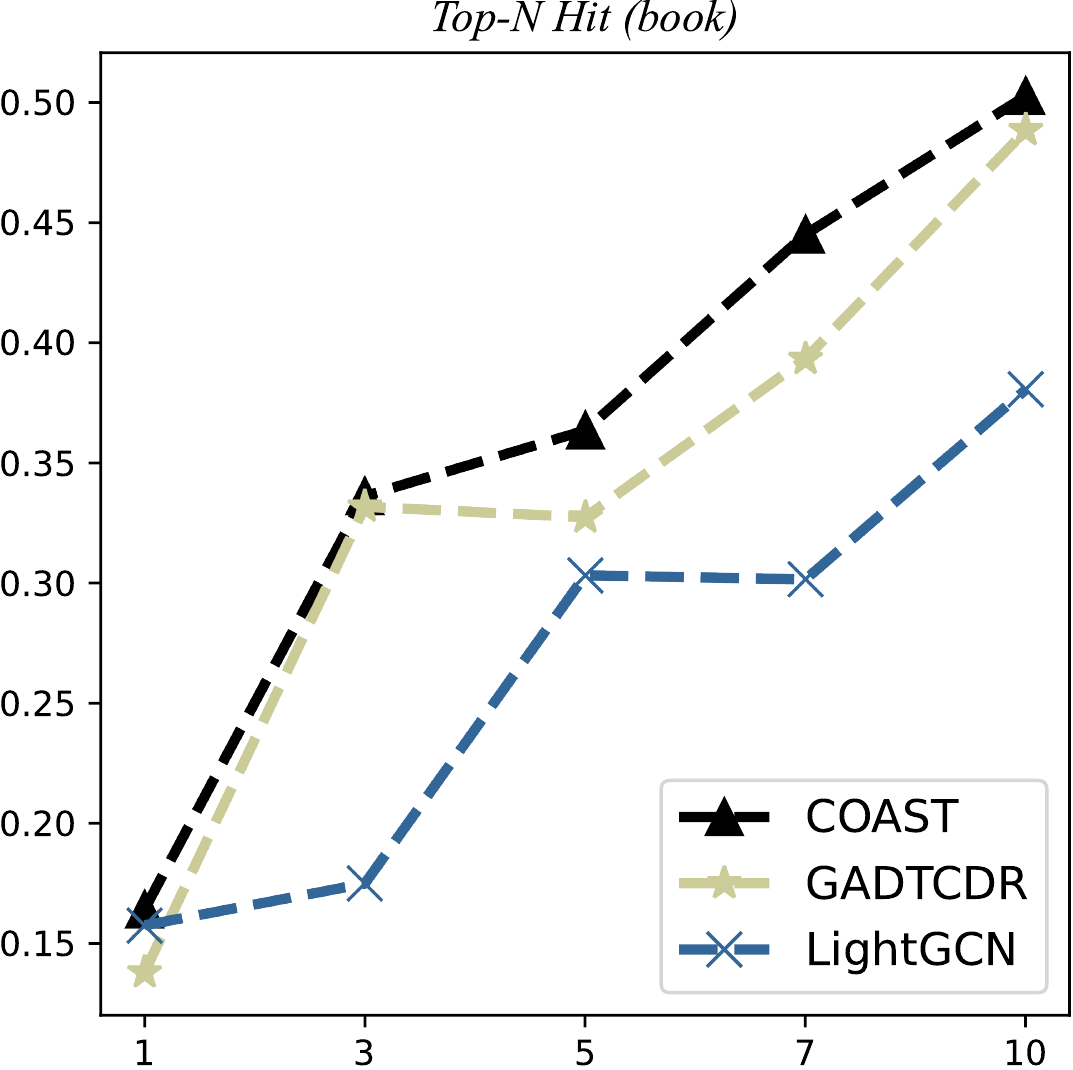}
\end{minipage}%
}%
\subfigure[NDCG of Douban-book.]{
\begin{minipage}[t]{0.45\linewidth}
\centering
\includegraphics[width=\linewidth,height=0.75\linewidth]{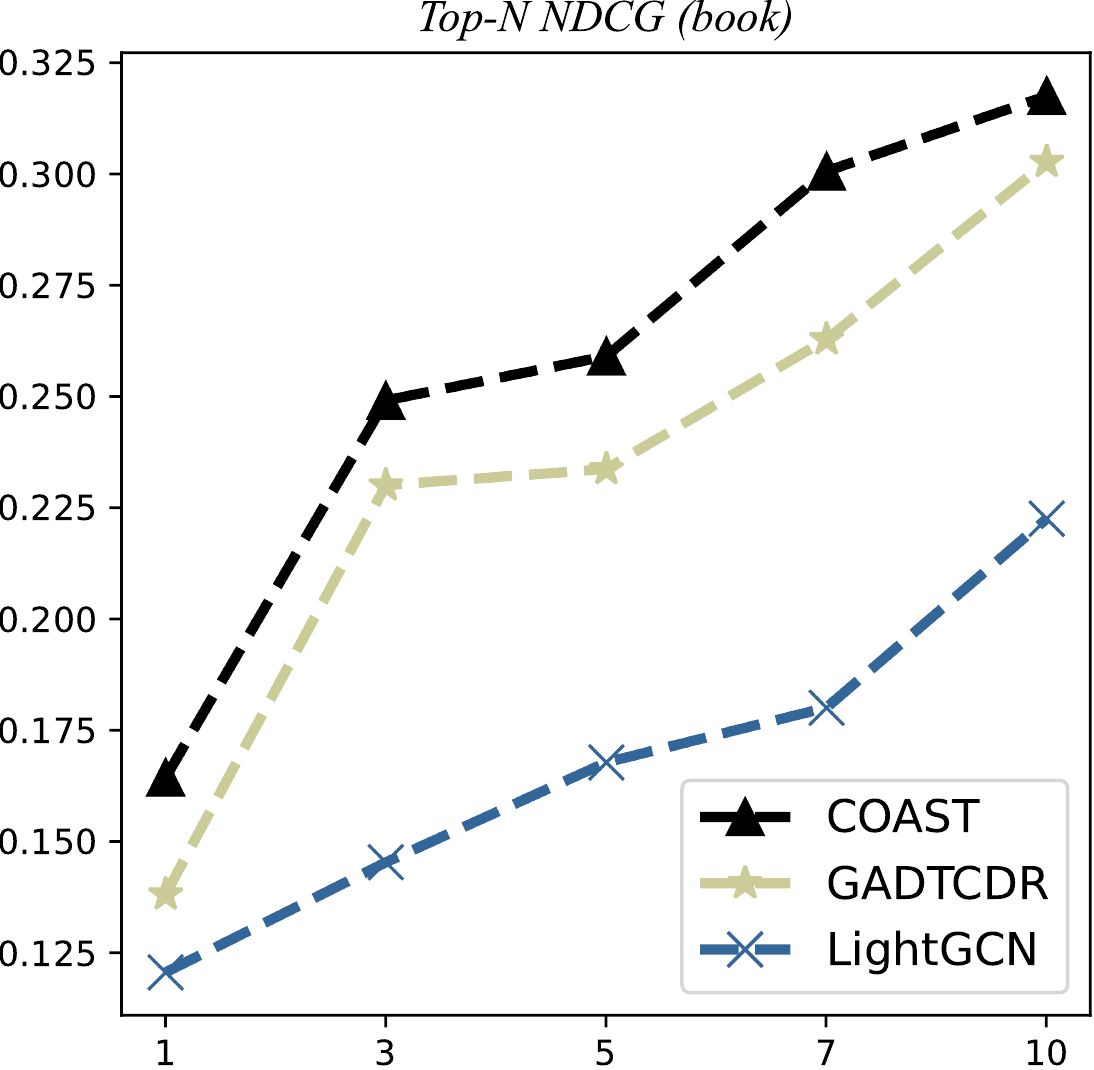}
\end{minipage}%
}%
\centering
\caption{Top-N performance.}
\label{topn}
\end{figure}
We also examine the performance of \textit{COAST} as well as the most competitive algorithms in single-domain, cross-domain baselines, i.e., LightGCN, GADTCDR, on different recommendation list lengths, as shown in Figure ~\ref{topn}.

Obviously, the performance of all algorithms increases as the recommendation list grows, because the longer the list, the higher the fault tolerance. Meanwhile, compared with the LightGCN and GADTCDR algorithms, our algorithm achieves the best performance in all scenarios, especially in the difficult $N=3$ scenario with the greatest improvement, which shows our superiority.

\subsubsection{Overlap Ratio M}
To investigate the robustness of our model, we experiment with scaling the number of overlapping users.

Table~\ref{overlap} reports the recommendation performance of \textit{COAST}, GADTCDR trained on corresponding cross-domain scenarios with overlapping users of 25\%, 50\%, 75\%, and 100\%, respectively. From Table~\ref{overlap}, we have the following observations. (1) With the increase of the overlapping user training ratio,the recommendation performance of all algorithms steadily improves, which demonstrates that overlapping ratio is effective to enhance the correlation across domains. (2) Our model shows robust performance to make recommendations for both domains than the strongest baseline GADTCDR, even with only 25\% user overlap. This is attributed to the unified graph message passing mechanism and user interest alignment, which enable the model to perceive the cross-domain similarity between entities and ensure consistent interests across views. (3) Further, we observe that the overlap ratio has little improvement on 75\%$\rightarrow$100\% than 25\% $\rightarrow$ 50\%, as the absolute number of overlapping users is large enough to ensure basic knowledge transfer.
\begin{table}
\centering
\caption{Overlap ratio test.}
\label{overlap}
\resizebox{0.48\textwidth}{!}{
\begin{tabular}{cc|cc|cc||cc|cc} 
\toprule
\multirow{3}{*}{Task}  & \multirow{3}{*}{Ratio} & \multicolumn{4}{c||}{COAST}                                & \multicolumn{4}{c}{GADTCDR}                               \\ 
\cline{3-10}
                       &                        & \multicolumn{2}{c|}{source} & \multicolumn{2}{c||}{target} & \multicolumn{2}{c|}{source} & \multicolumn{2}{c}{target}  \\ 
\cline{3-10}
                       &                        & Hit    & NDCG               & Hit    & NDCG                & Hit    & NDCG               & Hit    & NDCG               \\ 
\hline
\multirow{4}{*}{Task1} & 25\%                   & 0.6606 & 0.4013             & 0.4531 & 0.2735              & 0.6067 & 0.3447             & 0.3933 & 0.2412             \\
                       & 50\%                   & 0.6824 & 0.4242             & 0.4801 & 0.2976              & 0.6193 & 0.3711             & 0.4402 & 0.2709             \\
                       & 75\%                   & 0.6831 & 0.4132             & 0.5019 & 0.3078              & 0.6263 & 0.3766             & 0.4474 & 0.2889             \\
                       & 100\%                  & 0.6905 & 0.4271             & 0.5052 & 0.3174              & 0.6817 & 0.4205             & 0.4882 & 0.3026             \\ 
\hline
\multirow{4}{*}{Task2} & 25\%                   & 0.6875 & 0.4188             & 0.4055 & 0.2238              & 0.5997 & 0.3527             & 0.2805 & 0.1512             \\
                       & 50\%                   & 0.6881 & 0.4161             & 0.4372 & 0.2445              & 0.6186 & 0.3601             & 0.3301 & 0.1297             \\
                       & 75\%                   & 0.6872 & 0.4143             & 0.4382 & 0.2469              & 0.6101 & 0.3712             & 0.3445 & 0.1808             \\
                       & 100\%                  & 0.6938 & 0.4292             & 0.4497 & 0.2515              & 0.6818 & 0.4276             & 0.4383 & 0.2498             \\ 
\hline
\multirow{4}{*}{Taks3} & 25\%                   & 0.4763 & 0.2958             & 0.3929 & 0.2148              & 0.4080 & 0.2484             & 0.2805 & 0.1512             \\
                       & 50\%                   & 0.4845 & 0.3081             & 0.3954 & 0.2181              & 0.4338 & 0.2698             & 0.3367 & 0.1812             \\
                       & 75\%                   & 0.5014 & 0.3115             & 0.4192 & 0.2293              & 0.4350 & 0.2619             & 0.3375 & 0.1891             \\
                       & 100\%                  & 0.5138 & 0.3293             & 0.4688 & 0.2712              & 0.4492 & 0.2761             & 0.3571 & 0.1933             \\ 
\hline
\multirow{4}{*}{Task4} & 25\%                     &   0.6453     &   0.3883                &  0.5240       &    0.3454                 & 0.6380 & 0.3713             & 0.5037 & 0.3000             \\
                       & 50\%                   &   0.6550      &   0.3912                 &     0.5236   &    0.2998                 & 0.6470 & 0.3863             & 0.5069 & 0.2916             \\
                       & 75\%                  &    0.6590     &  0.3945                  &    0.5439    &    0.3242                 & 0.6493 & 0.3874             & 0.5010 & 0.2922             \\
                       & 100\%                  &   0.6769     &                  0.4073  &    0.5503    &  0.3195                   & 0.6654 & 0.4055             & 0.5173 & 0.2907             \\
\bottomrule
\end{tabular}
}
\end{table}

\subsubsection{Ablation Studies}
We further compare \textit{COAST} with several ablation variants to demonstrate the effectiveness and advancement of different sub-modules. For fairness, other settings are kept unchanged except for the specified ablation module.
\begin{itemize}[leftmargin=12pt]
    \item \textbf{COAST-NF:} This variant uses only explicit interactions.
    \item \textbf{COAST-NS:} Instead of constructing cross-domain heterogeneous graphs, each domain trains representations separately.
    \item \textbf{COAST-NM:} No user-item interaction in section~\ref{inters}.
    \item \textbf{COAST-NU:} This variant is not subject to user-user consistency.
    \item \textbf{COAST-NI:} This variant is not subject to user-item consistency.
\end{itemize}
\begin{figure}
\centering
\subfigure[Hit@10 of Douban-movie.]{
\begin{minipage}[t]{0.45\linewidth}
\centering
\includegraphics[width=\linewidth,height=0.80\linewidth]{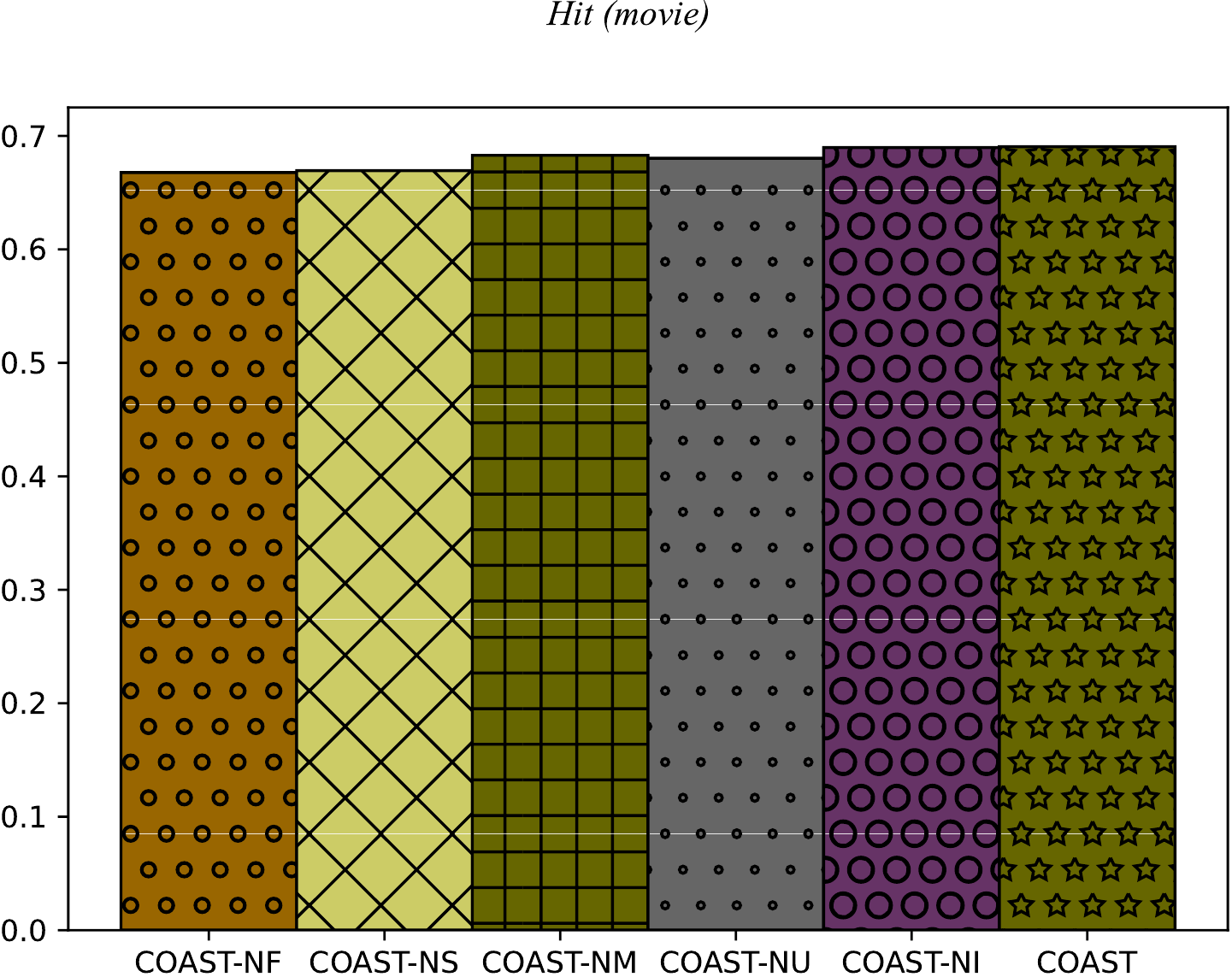}
\end{minipage}%
}%
\subfigure[NDCG@10 of Douban-movie.]{
\begin{minipage}[t]{0.45\linewidth}
\centering
\includegraphics[width=\linewidth,height=0.80\linewidth]{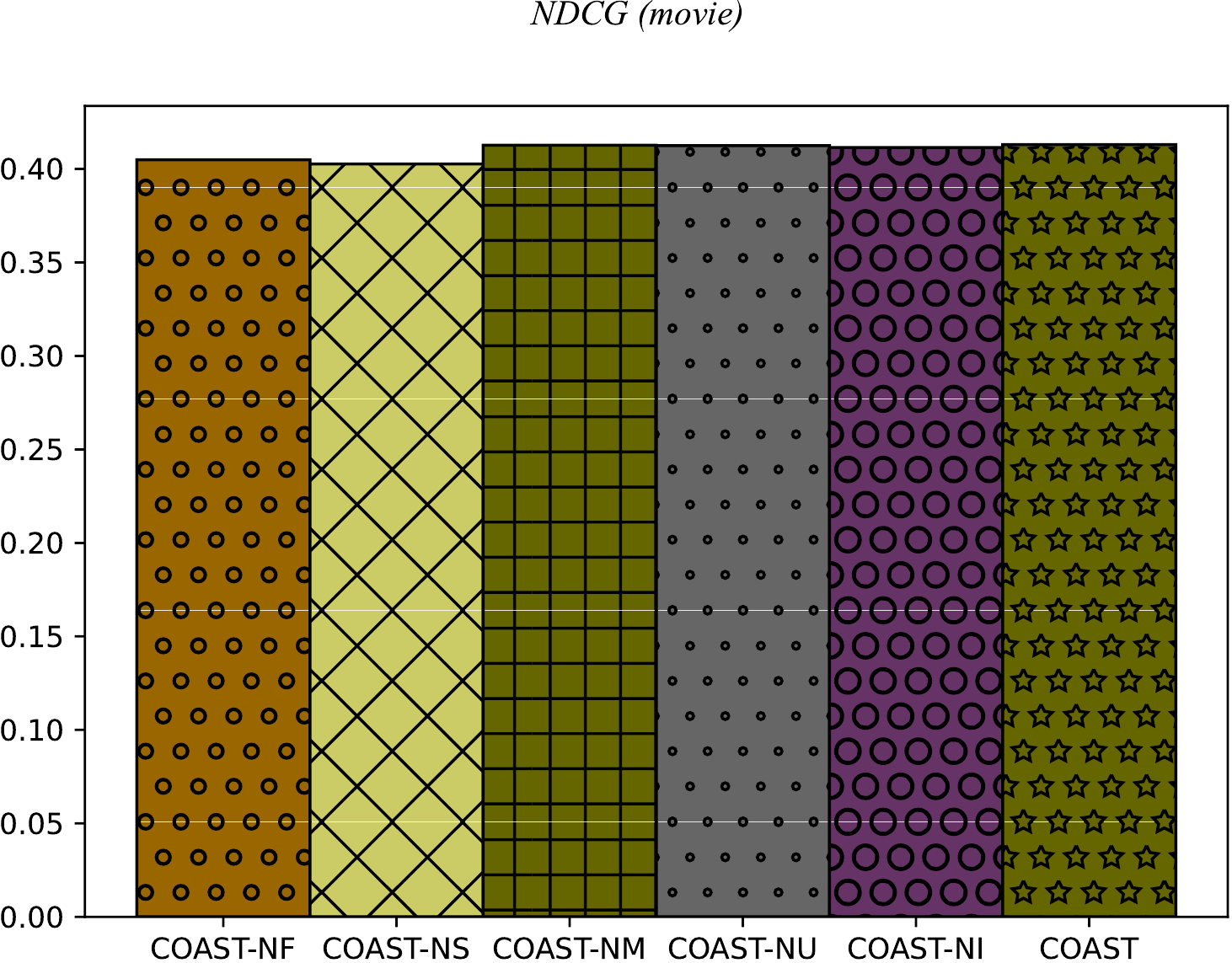}
\end{minipage}%
}%
\vskip\baselineskip
\subfigure[Hit@10 of Douban-book.]{
\begin{minipage}[t]{0.45\linewidth}
\centering
\includegraphics[width=\linewidth,height=0.80\linewidth]{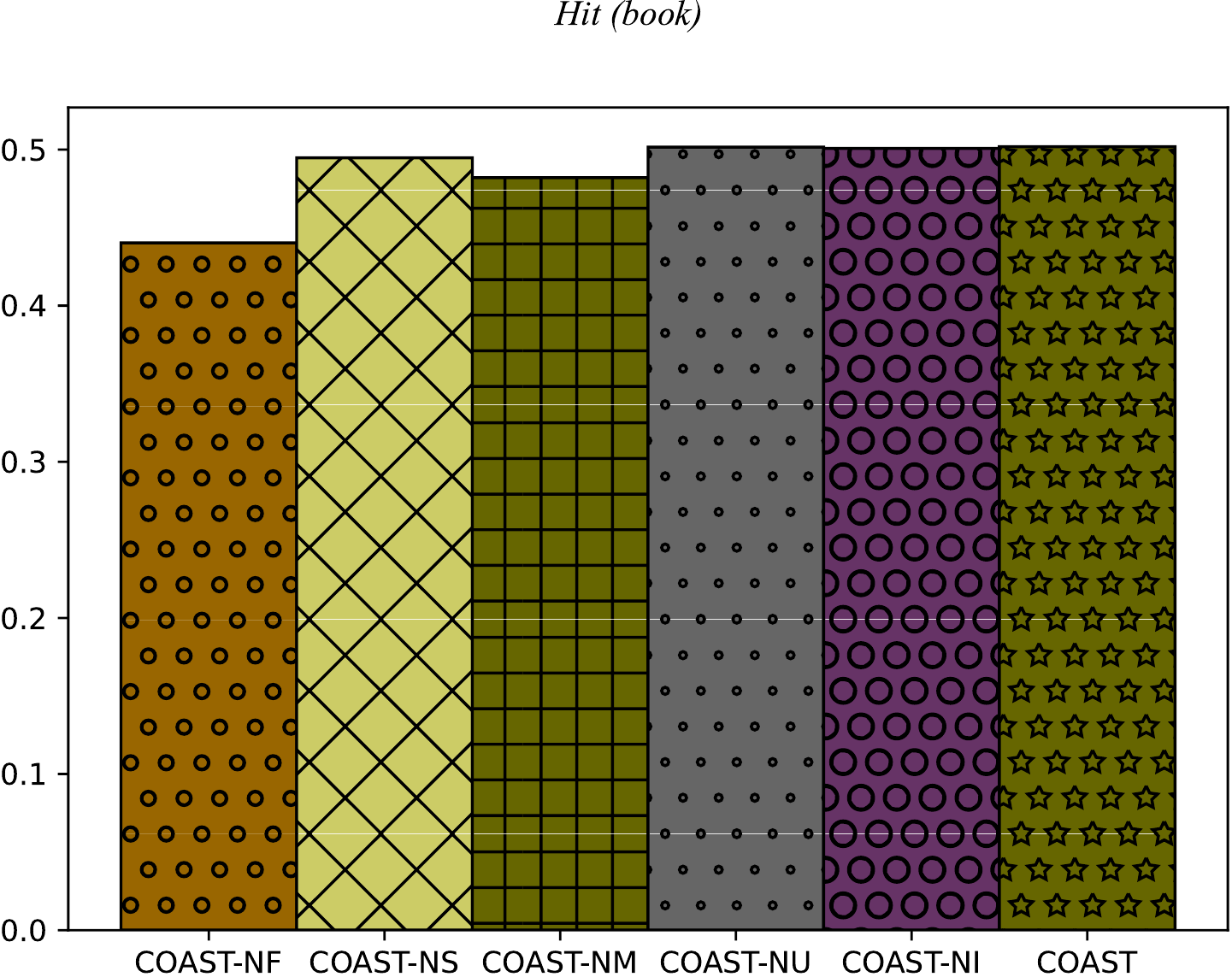}
\end{minipage}%
}%
\subfigure[NDCG@10 of Douban-book.]{
\begin{minipage}[t]{0.45\linewidth}
\centering
\includegraphics[width=\linewidth,height=0.80\linewidth]{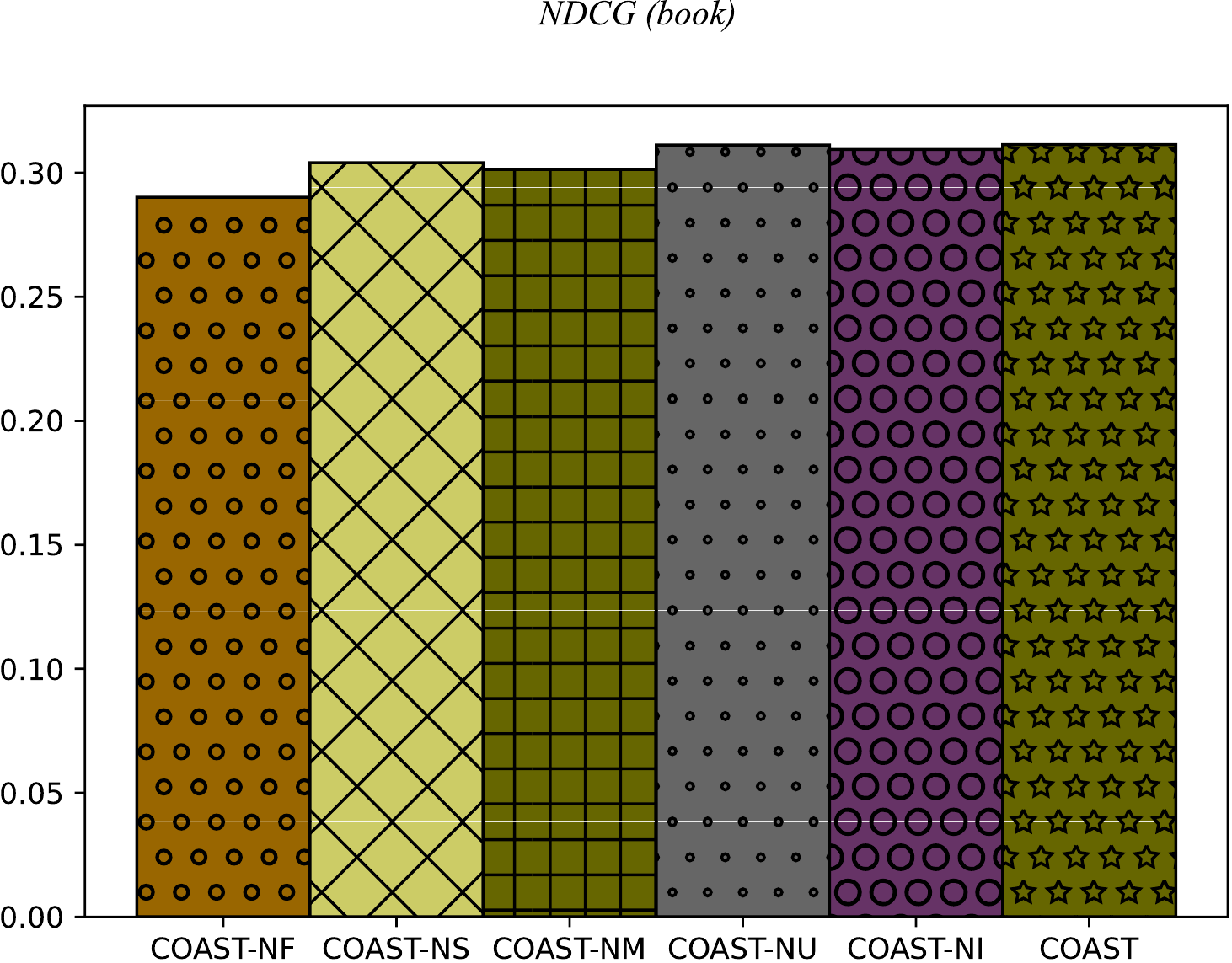}
\end{minipage}%
}%
\centering
\caption{Ablation studies.}
\label{aba}
\end{figure}

As reported in Figure~\ref{aba}, COAST-NF has the worst performance but is still stronger than the vast majority of baselines (except for GADTCDR), illustrating that our model structure is able to mine structural similarities from explicit data. 
Regarding COAST-NS and COAST-NM, as ablations of the cross-domain graph module, both decrease compared with \textit{COAST}. The former cannot capture cross-domain similarity due to the isolation of user-item cross-domain interactions at the graph structure level, while the latter is insufficient to characterize the collaborative filtering relationship due to ignoring the collaborative signal of user-item.
Meanwhile, with the same structure, \textit{COAST} improve over COAST-NU, COAST-NI. This demonstrates that using user interest alignment as a constraint can not only effectively prevent overfitting, but also, as a fine-grained knowledge utilization, significantly enhance the generalization of user representations across domains. From a deeper perspective, contrastive learning and gradient alignment leverage the potential unsupervised signals and semantic features in the data, which greatly facilitates the extraction of domain-invariant features.
In general, each submodule of \textit{COAST} plays an indispensable role and contributes significantly to the model performance.

\subsubsection{Hyper-testing}
In this subsection, we present the tuning of several key hyper-parameters in our framework.

\noindent\textbf{Embedding size \textit{D}.}
Embedding size is one of the most important hyper-
parameters in deep learning and is closely related to model capacity~\cite{zhao2022multi}. To improve the performance of the proposed \textit{COAST}, we perform a hyper-parameter search on the embedding size.

\begin{figure}
\centering
\subfigure[Hit@10 of Douban-movie.]{
\begin{minipage}[t]{0.45\linewidth}
\centering
\includegraphics[width=\linewidth,height=0.7\linewidth]{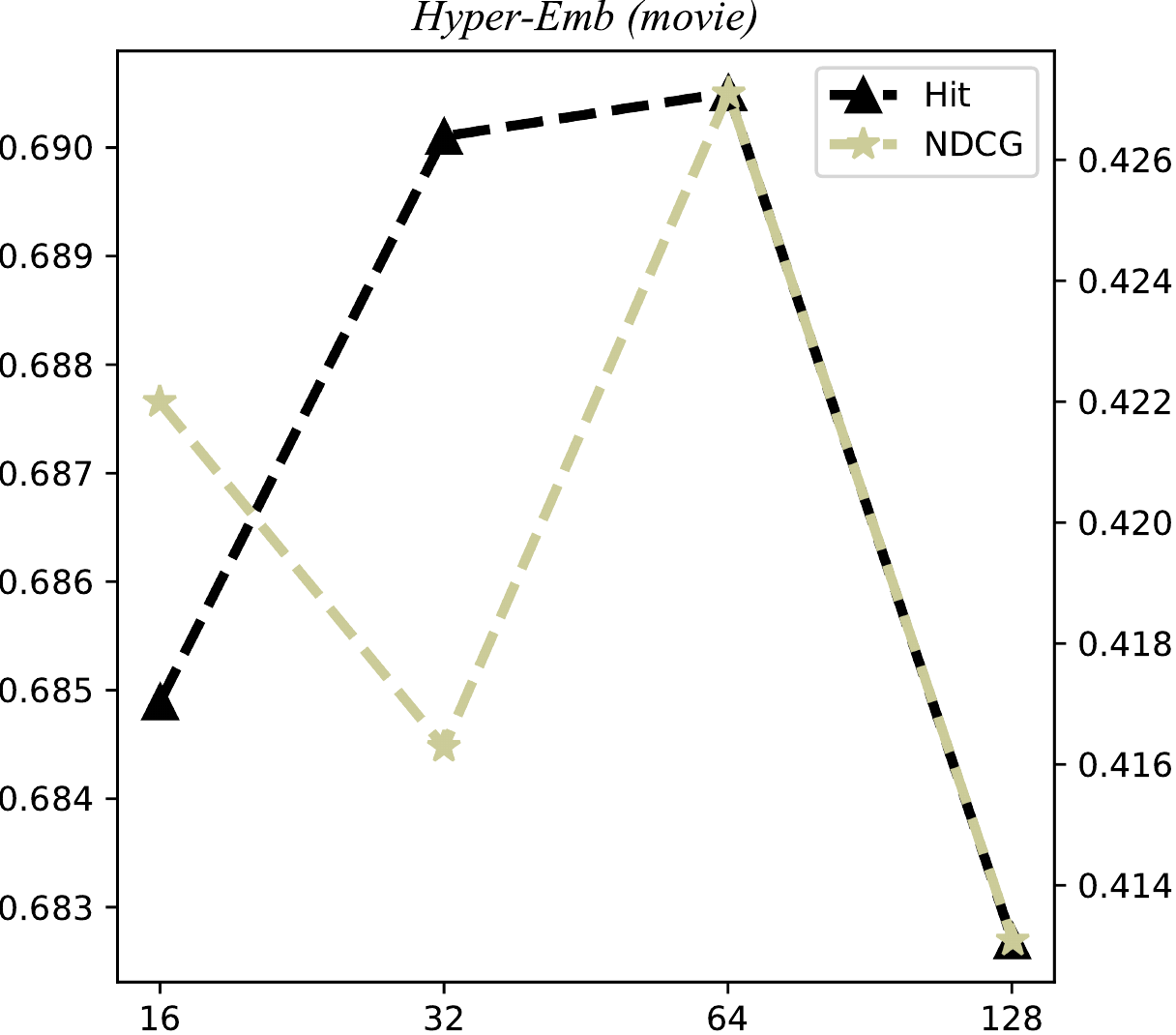}
\end{minipage}%
}%
\subfigure[Hit@10 of Douban-book.]{
\begin{minipage}[t]{0.45\linewidth}
\centering
\includegraphics[width=\linewidth,height=0.7\linewidth]{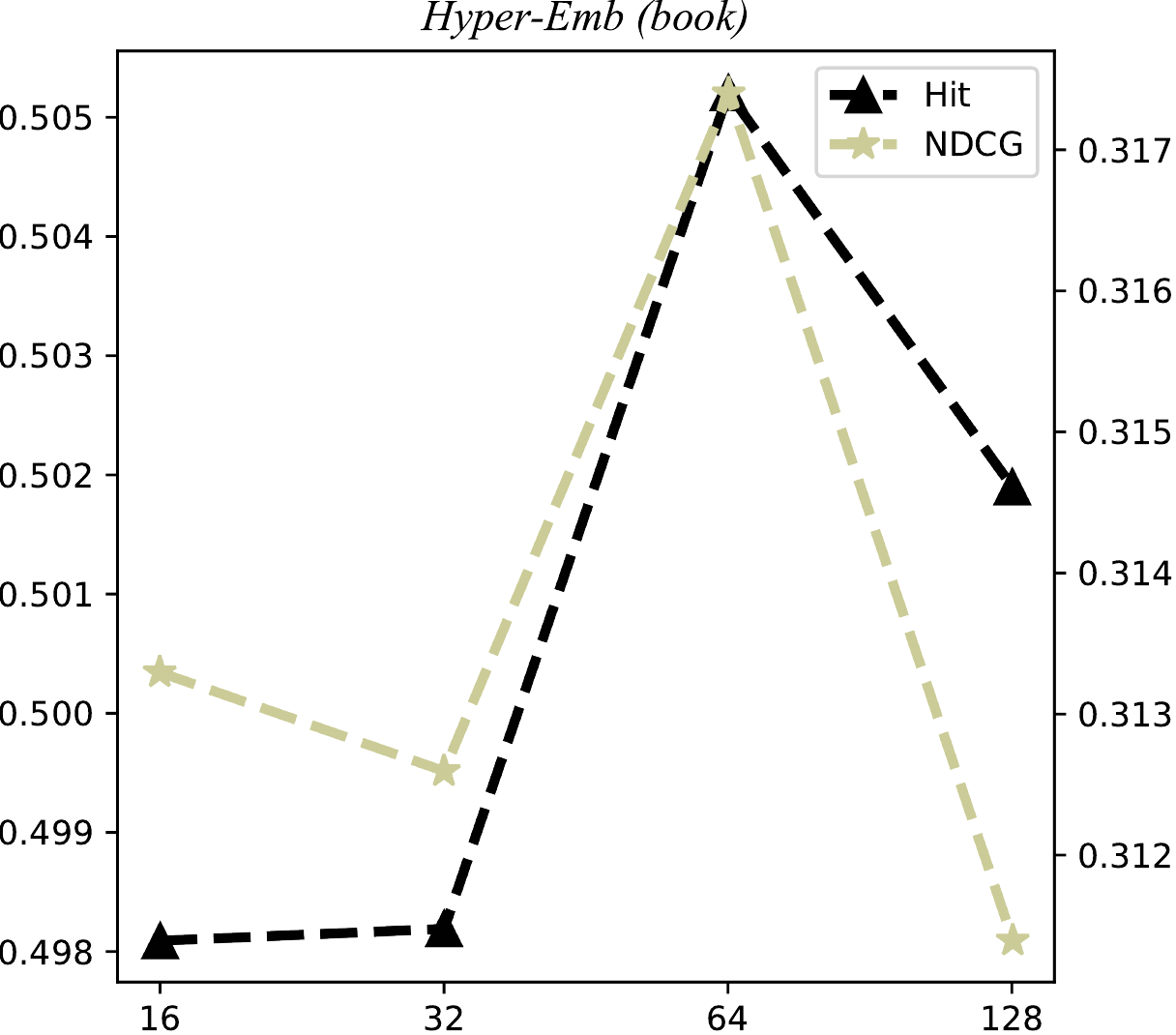}
\end{minipage}%
}%
\centering
\caption{The impact of $D$.}
\label{emb}
\end{figure}

As shown in Figure~\ref{emb}, our algorithm performs best when $D=64$ on any metrics. The larger the embedding size, the more expressive the model is, but too high embedding size will slow down the convergence speed and lead to overfitting. 
In consequence, we choose $D=64$ as the embedding size in \textit{COAST}.

\noindent\textbf{Number of Interests \textit{K}.}
In section~\ref{uu}, we constrain users' contrasting views to belong to the same cluster center. In view of this, we perform a test on the number of interest cluster centers $K$.
\begin{figure}
\centering
\subfigure[Hit@10 of Douban-movie.]{
\begin{minipage}[t]{0.45\linewidth}
\centering
\includegraphics[width=\linewidth,height=0.7\linewidth]{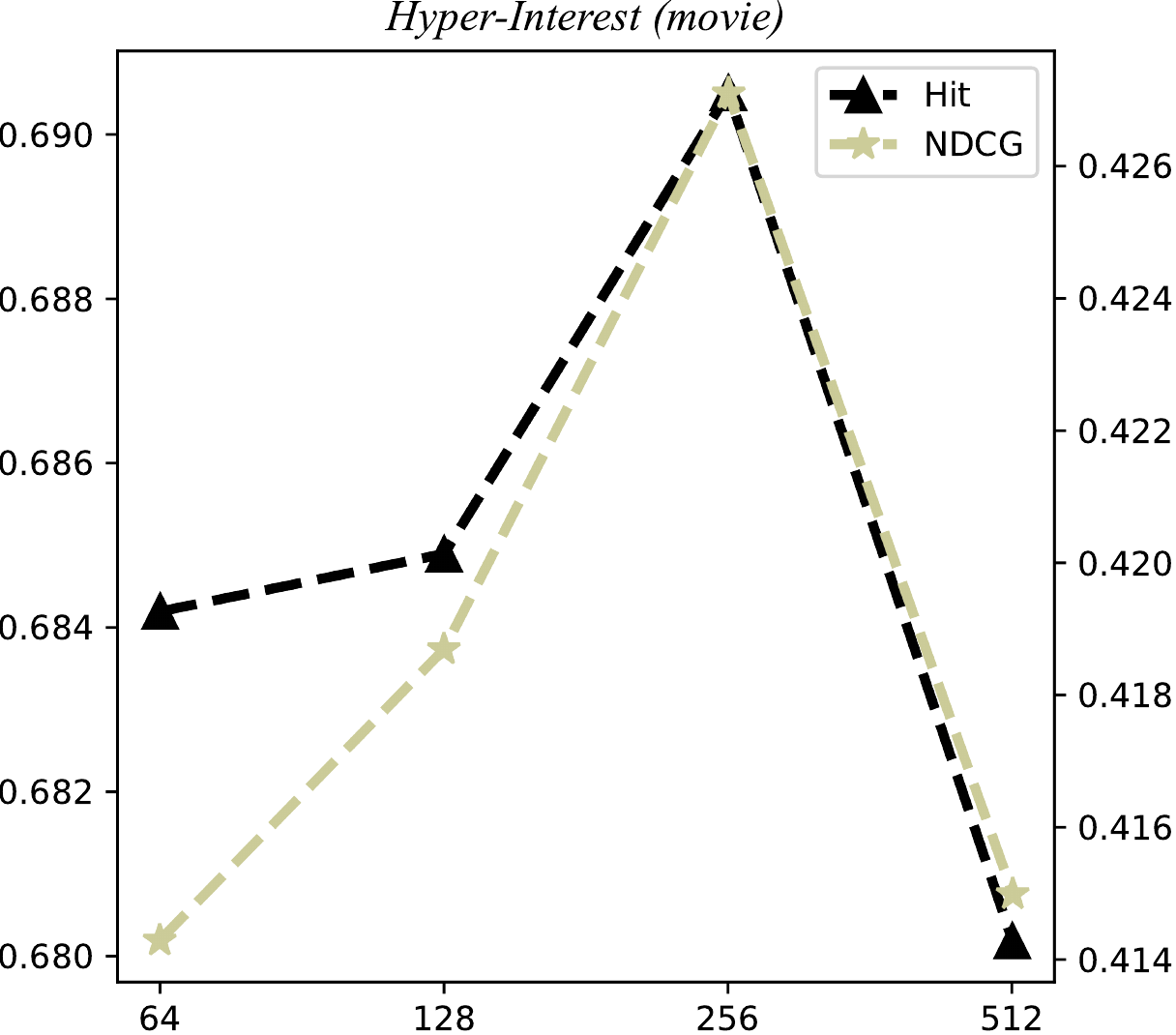}
\end{minipage}%
}%
\subfigure[Hit@10 of Douban-book.]{
\begin{minipage}[t]{0.45\linewidth}
\centering
\includegraphics[width=\linewidth,height=0.7\linewidth]{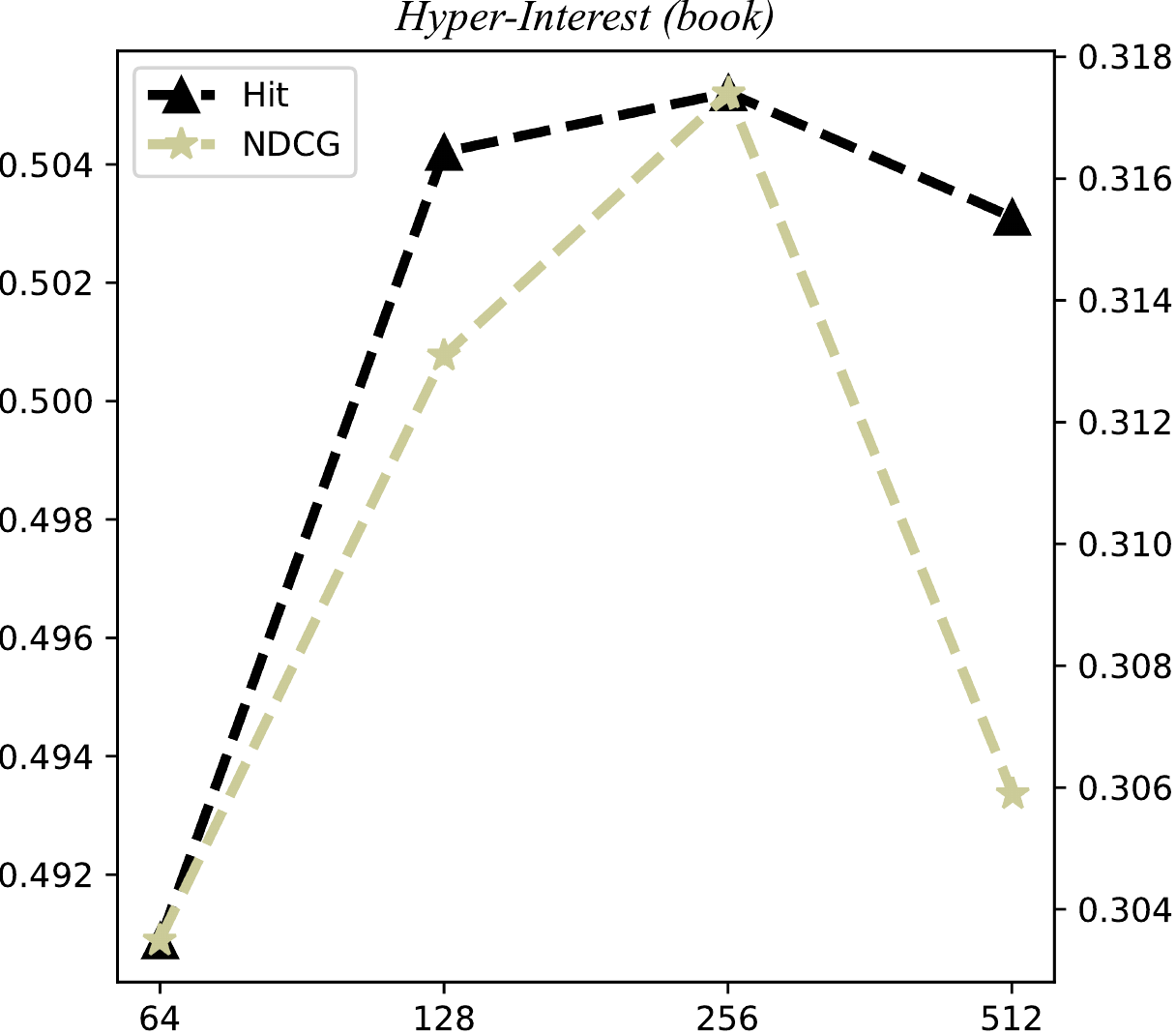}
\end{minipage}%
}%
\centering
\caption{The impact of $K$.}
\label{interest}
\end{figure}

As shown in Figure~\ref{interest}, our model is sensitive to $K$. We argue that this phenomenon arises because $K$ represents an abstract interest center rather than a concrete interest. Meanwhile, we propose that higher $K$ can be chosen to characterize the distribution of user interests when the number of items and users increases. This is intuitive, as the number of users increases, the interests will obviously become more diverse. Consequently, we choose $K=256$.

\noindent\textbf{Consistency weight $\lambda_2$.}
The consistency weight $\lambda_2$ is a trade-off between task interest and user interest alignment. To improve the recommendation effect, we have tuned it.
\begin{figure}
\centering
\subfigure[Hit@10 of Douban-movie.]{
\begin{minipage}[t]{0.45\linewidth}
\centering
\includegraphics[width=\linewidth,height=0.7\linewidth]{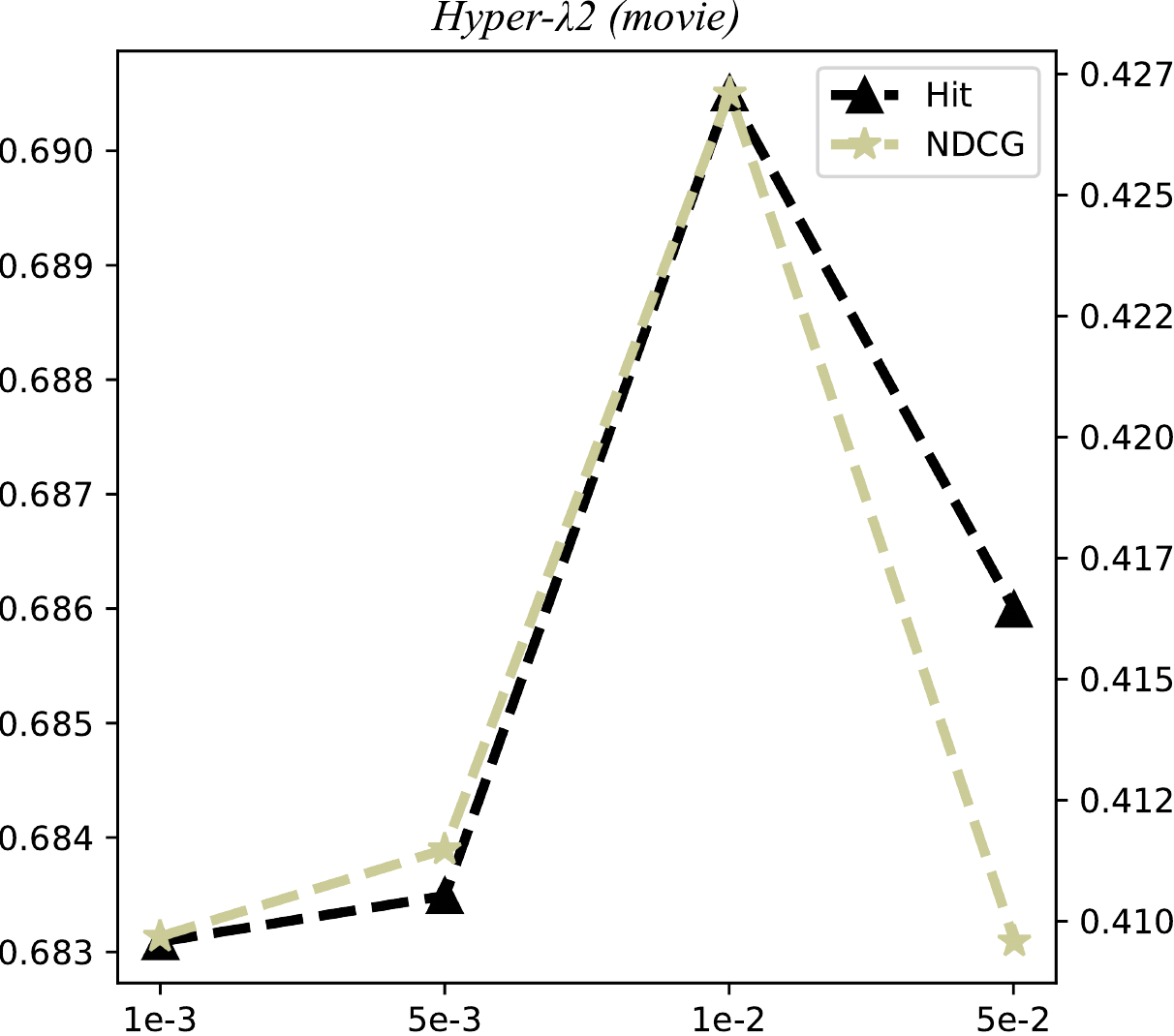}
\end{minipage}%
}%
\subfigure[Hit@10 of Douban-book.]{
\begin{minipage}[t]{0.45\linewidth}
\centering
\includegraphics[width=\linewidth,height=0.7\linewidth]{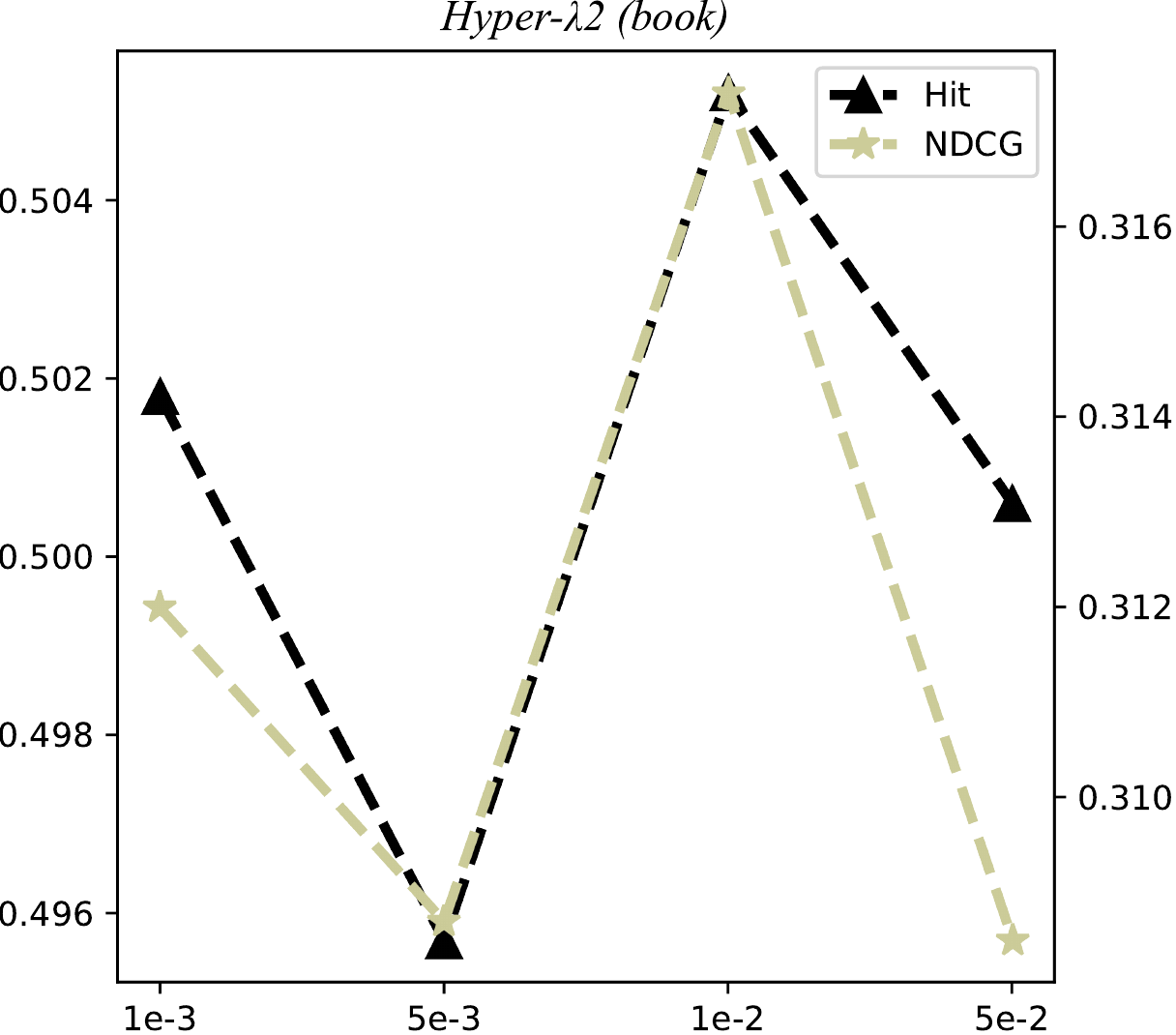}
\end{minipage}%
}%
\centering
\caption{The impact of $\lambda_2$.}
\label{win}
\end{figure}

The larger $\lambda_2$ is, the stronger the constraint on user interest consistency is, but hinders domain-specific user representation, thereby impairing recommendation performance on that domain. Conversely, the smaller $\lambda_2$ is, our model will degenerate into a general representation combination model, which cannot solve user interest alignment. Experimentally, we set $\lambda_{2}=0.01$.

\section{Conclusion}
In this work, we propose the \textit{COAST} framework, which aims to improve model performance in dual cross-domain recommendation scenarios. This work represents an attempt to leverage rich content information and user interest alignment for bidirectional knowledge transfer. Specifically, we model the interaction of users and items in two domains as a unified cross-domain heterogeneous graph, and improve the message passing mechanism of graph convolution to capture the cross-domain similarity of users and items. Further, we utilize contrastive learning and gradient alignment to constrain overlapping user interest alignment from both user-user and user-item perspectives. Overall, our solution has several advantages. First, at the data level, our task is constructed on data sets with partial user overlap and exploits both explicit and implicit information, which has a wider range of application scenarios. Second, at the algorithm level, we learn better representations from high-order cross-domain similarity and user interest alignment compared to previous plain combinations. Finally, at the experimental level, we conduct extensive experiments, all of which demonstrate the state-of-the-art and superiority of our model.

There are still several limitations of our study for future work. First, how to extend our work to more complex scenarios, such as the case of overlapping items or multi-domain recommendation.
Second, how to integrate data from other modalities or integrate more complex interactions, such as images, attribute nodes, in  feature extraction module.
Finally, we should validate the robustness of \textit{COAST} on more large cross-domain recommendation data sets.
\bibliographystyle{ACM-Reference-Format}
\bibliography{main}

\clearpage
\appendix

\section{Important Notations}\label{A}
\begin{table}[!h]
\centering
\caption{Mathematical Notation}\label{math}
\begin{tabular}{c|c} 
\toprule
\textbf{Symbol}                                                & \textbf{Notation}                              \\ 
\hline
$\mathcal{S},\mathcal{T}$                                      & source/target domain                           \\
$\mathcal{U}$                                                  & user set                                       \\
$\mathcal{V}$                                                  & item set                                       \\
$\mathcal{A}$                                                  & user-item interaction matrix                   \\
$\mathcal{X},\mathcal{H}$                                      & features before/after preprocessing            \\
$\mathcal{G}=(\mathcal{U},\mathcal{V},\mathcal{E},\mathcal{H})$ & heterogeneous graph of user-item interactions  \\
$N$                                                            & neighbors set                                  \\
$m$                                                            & message passing function                       \\
$K$                                                            & the total number of interests in the user set  \\
$g$                                                            & gradient calculation                           \\
$y$                                                            & whether the user clicked on the item                           \\
\bottomrule
\end{tabular}
\end{table}

\section{Algorithm}\label{B}
\begin{algorithm}[!h] 
\begin{small}

\caption{The Algorithm of \textit{COAST}} 
\label{alg1} 
\begin{algorithmic}[1] 
\REQUIRE Interaction matrix ${\mathcal{A_S}}$,${\mathcal{A_T}}$,$\mathcal{X_{S}}$,$\mathcal{X_T}$;
\ENSURE Parameters $\Theta$;
\STATE Random initialize model parameters $\Theta$, 
\STATE Data preprocessing $e^{u} \in \mathcal{H_U}$, $e^{v}_{\mathcal{S}} \in {\mathcal{H_S}}$, $e^{v}_{\mathcal{T}} \in \mathcal{H_T}$
\STATE Graph construction $\mathcal{G}=\mathcal{(U, V, E, H)}$
\WHILE{not converged}
\STATE Sample a batch of training data
\STATE Graph propagation, getting $e^{u}$, $e^{v}$
\FOR{$u \in \mathcal{U}_{o}$}
\STATE User-User interest alignment $\mathcal{L}_{U,U}$
\STATE User-Item interest alignment $\mathcal{L}_{U,I}$
\ENDFOR
\STATE Supervise loss $\mathcal{L}_{s}$
\STATE Joint optimization $\mathcal{L}$
\STATE Update the parameters
\ENDWHILE
\RETURN Parameters $\Theta$
\end{algorithmic}
\end{small}
\end{algorithm}

\end{document}